\documentclass[a4paper,aps,amsmath,onecolumn,amssymb,nopacs,10pt]{revtex4-2}
\usepackage{epsfig,bm}
\usepackage{graphicx,float,subcaption}
\usepackage{amsmath,amssymb,psfrag,textcomp}
\usepackage{amsthm,upgreek} 
\usepackage{booktabs}
\usepackage{siunitx}
\usepackage[usenames,dvipsnames,svgnames,table]{xcolor}
\usepackage{hyperref}
\hypersetup{colorlinks=true, linkcolor=NavyBlue, citecolor=PineGreen,urlcolor=cyan}

\newcommand{\bracket}[1]{\left\langle #1\right\rangle}

\newcommand{\be}{\begin{equation}}
\newcommand{\ee}{\end{equation}}
\newcommand{\bd}{\begin{displaymath}}
\newcommand{\ed}{\end{displaymath}}

\newcommand{\beeq}[1] {\begin{equation}\begin{split}#1\end{split}\end{equation}}

\begin{document}
\title{Beyond Whittle: exact finite-time multispectral statistics from a single Brownian trajectory in a harmonic trap}
\author{Isaac P\'erez Castillo}
\email{iperez@izt.uam.mx}
\affiliation{Departamento de F\'isica, Universidad Autónoma Metropolitana-Iztapalapa, San Rafael Atlixco 186, Ciudad de México 09340, Mexico}
\affiliation{Instituto de Ciencias Físicas, Universidad Nacional Autónoma de México, Av. Universidad s/n, Col. Chamilpa, Cuernavaca, 62210, Mexico}
\author{François Leyvraz}
\affiliation{Instituto de Ciencias Físicas, Universidad Nacional Autónoma de México, Av. Universidad s/n, Col. Chamilpa, Cuernavaca, 62210, Mexico}
\author{Miguel Eduardo G\'omez Quintanar}
\author{Andr\'es \'Alvarez Ballesteros}
\affiliation{Departamento de F\'isica, Universidad Autónoma Metropolitana-Iztapalapa, San Rafael Atlixco 186, Ciudad de México 09340, Mexico}
\begin{abstract}
Power spectral densities are often interpreted through ensemble averages and long-time asymptotics. In many experiments, however, only a single finite record is available, so spectral estimators remain broadly distributed and the usual independence assumptions across frequencies need not hold. Here we develop an exact finite-$T$ multispectral theory for an overdamped Brownian particle in a harmonic trap. For a collection of frequencies $\{\omega_i\}$, we obtain an exact characterization of the joint law of the finite-time estimators $\{S(\omega_i,T)\}$, together with a covariance-explicit Gaussian representation for the associated Fourier projections. This representation makes the observation-window-induced inter-frequency correlations explicit and shows how they vanish as $T\to\infty$, thereby recovering the asymptotic Whittle picture. We then use this structure to formulate a hierarchy of spectral likelihoods for inference from a single trajectory, ranging from the factorized Whittle approximation to blockwise covariance-aware approximations in frequency space. Monte Carlo simulations validate the finite-time theory and quantify the effect of neglected cross-frequency correlations on single-trajectory estimates of the trap parameters. Our results provide a controlled finite-time benchmark for spectral inference beyond the asymptotic regime.
\end{abstract}

\maketitle
%\tableofcontents

%%%%%%%%%%%%%%%%%%%%%%%%%%%%%%%%%%%%%%%%%%%%%%%%%%%
\section{Introduction}
%%%%%%%%%%%%%%%%%%%%%%%%%%%%%%%%%%%%%%%%%%%%%%%%%%%
The power spectral density (PSD) of a time series $X(t)$ characterizes how the variance (or ``power'') of the signal is distributed across frequencies and often encodes key dynamical information about the underlying process. For this reason, spectral analysis is a standard tool across a wide range of disciplines, including musical signals \cite{voss19751,hennig2011nature}, climate data \cite{weber2001}, inter-event statistics in seismology \cite{Sornette1989}, and the calibration of optical tweezers \cite{BergSorensenFlyvbjerg2004,WongHalvorsen2006,PerezGarcia2023,gieseler2021optical,volpe2022roadmap}, among many others.

In the simplest setting, the PSD of a stationary signal $X(t)$ can be defined as the Fourier transform of its autocorrelation function or equivalently as a long-time and ensemble limit of the finite-window Fourier amplitude \cite{GrenanderRosenblatt1957,Priestley1981,Brillinger1981,BrockwellDavis1991}, that is
\beeq{
\mu(\omega)=\lim_{T\to\infty} \frac{1}{T}\mathbb{E}_{\{X(t)\}}\left[\left|\int_0^T dt e^{i \omega t} X(t) \right|^2\right]\,,
}
where $\mathbb{E}_{\{X(t)\}}[\cdots]$ denotes ensemble average with respect to the underlying stochastic process that produces the signal $X(t)$ and $|\cdots|$ the modulus of a complex number. In many practical situations, however, one only has access  to a small number of realizations and to finite acquisition times, so neither the ensemble average nor the limit $T\to\infty$ can be faithfully approximated. This is the case, for example, in measurements on \textit{in vivo} systems \cite{norregaard2017manipulation}, in climate-related records \cite{kemp2015maximum}, in some cases of financial time series \cite{Lemperiere2011}, and in fluctuation-based calibration protocols where finite integration time and sampling rate directly affect the observed spectrum \cite{WongHalvorsen2006,PerezGarcia2023}.

These considerations motivate the study of spectral analysis in the regime of finite acquisition times and single-trajectory data. Recent work has clarified the statistical structure of single-trajectory spectral estimators well beyond their mean PSD, including exact single-frequency distributions, universal noise-to-signal properties for Gaussian processes, and explicit frequency-frequency correlations \cite{krapf2018power,krapf2019spectral,sposini2019single,squarcini2022noise,squarcini2022frequency}. In parallel, the statistics literature has shown that standard Whittle-type frequency-domain likelihoods can exhibit substantial finite-sample distortions due to leakage, aliasing, and boundary effects, which has motivated debiased and boundary-corrected variants \cite{Whittle1953,sykulski2019debiased,subbarao2021reconciling,das2021spectral}. The present work lies at the intersection of these two directions.

Our goal is to derive the exact finite-$T$ multispectral statistics of an overdamped Brownian particle in a harmonic trap, equivalently an Ornstein--Uhlenbeck (OU) process, from a single observed trajectory. More precisely, for a collection of frequencies $\{\omega_i\}$ we obtain an exact characterization of the joint law of the finite-time estimators $\{S(\omega_i,T)\}$ through its multivariate Laplace transform, together with a covariance-explicit Gaussian representation for the associated Fourier projections. This makes the observation-window-induced inter-frequency couplings explicit at finite $T$ and shows how the asymptotic Whittle picture emerges only when these couplings vanish. In this sense, the paper goes beyond the existing single-frequency theory and beyond correlation diagnostics alone: it provides an exact finite-time multispectral framework that can be used directly to construct frequency-domain likelihoods and to assess the error introduced by Whittle-type factorizations.

This multispectral perspective is relevant, from a practical viewpoint, because spectral characterization rarely relies on a single frequency. One typically evaluates estimators on a set of frequencies in order to resolve characteristic time scales and to fit parametric models. At finite observation time, however, spectral estimates at different frequencies are generally not independent, since the observation window couples Fourier modes and induces nontrivial inter-frequency correlations. Factorized likelihoods then overstate the effective amount of independent information carried by the spectrum. The consequence is not merely a possible shift in point estimates, but a systematic mis-specification of the finite-$T$ likelihood itself. For the Ornstein--Uhlenbeck model considered here, exact time-domain Gaussian likelihoods are in principle available, so the spectral problem is especially useful as a controlled setting in which to diagnose precisely what is lost when one replaces the exact finite-time dependence structure by asymptotic frequency-wise factorization.

Related work in machine learning has revisited frequency-domain likelihoods as a route to scalable probabilistic modeling of time series. A representative example is the Whittle Network construction of Yu \emph{et al.} \cite{yu2021whittle}, which enriches the factorized Whittle baseline by learning conditional-independence structure in the spectral domain. Our perspective is complementary and analytically grounded: for a canonical Gaussian dynamics we derive the exact finite-$T$ dependence pattern generated by time windowing, thereby providing a benchmark for when Whittle factorization is accurate, when it fails, and how it can be systematically refined.

This paper is organized as follows. Section~\ref{definitions} introduces the finite-time spectral estimators, the OU dynamics, and the notation that connects the continuous-time theory to the discretely sampled inference problem studied later. In Section~\ref{onefrequency} we derive the exact finite-$T$ probability law of the single-frequency spectral estimator, which serves to illustrate the method and to connect with earlier single-frequency results. Section~\ref{multiplefrequency} contains the main theoretical contribution: the finite-$T$ multispectral characterization across several frequencies, together with the induced inter-frequency correlations and a Gaussian representation in terms of Fourier coefficients. In Section~\ref{sec:inference} we use this structure to benchmark a hierarchy of frequency-domain pseudo-likelihoods, ranging from the standard Whittle approximation to blockwise covariance approximations, and to quantify the effect of neglected cross-frequency dependence on single-trajectory inference. We conclude in Section~\ref{conclusions} with a discussion and outlook; technical derivations are deferred to the appendices.

%%%%%%%%%%%%%%%%%%%%%%%%%%%%%%%%%%%%%%%%%%%%%%%%%%%
\section{Definitions}
\label{definitions}
%%%%%%%%%%%%%%%%%%%%%%%%%%%%%%%%%%%%%%%%%%%%%%%%%%%
We begin by fixing the finite-window spectral observables and the stochastic model used throughout this work. Given a realization $X(t)$ of a stochastic process observed over a finite interval $[0,T]$, a natural spectral estimator at angular frequency $\omega$ is
\beeq{
S(\omega,T)&=\frac{1}{T}\left|\int_0^T dt e^{i\omega t} X(t)\right|^2=\frac{1}{T}\int_0^T\int_0^T dt_1dt_2 \cos\big[\omega(t_1-t_2)\big] X(t_1)X(t_2)\,.
}
Thus $S(\omega,T)$ is the spectral observable associated with a single finite record. Since $X(t)$ is random, $S(\omega,T)$ is itself a random variable and fluctuates from one realization to another.

Its asymptotic mean defines what is commonly referred to as the PSD, namely
\beeq{
\mu(\omega)=\lim_{T\to\infty}\frac{1}{T}\mathbb{E}_{\{X(t)\}}\left[\left|\int_0^T dt e^{i \omega t}\, X(t) \right|^2\right]=\lim_{T\to\infty}\frac{1}{T}\int_0^T\int_0^T dt_1dt_2\cos\big[\omega(t_1-t_2)\big]\mathbb{E}_{\{X(t)\}}\left[X(t_1)X(t_2)\right]\,,
}
which, for centered stationary processes, is the standard time-domain form of the Wiener--Khinchin relation \cite{GrenanderRosenblatt1957,Priestley1981,Brillinger1981,BrockwellDavis1991}. At finite $T$, however, the mean alone is not sufficient: the relevant object is the full probability law of $S(\omega,T)$.

We therefore characterize $S(\omega,T)$ through its probability density function (PDF),
\beeq{
\rho_{\omega,T}(s) \equiv \mathbb{E}_{\{X(t)\}}\left[\delta\!\big(S(\omega,T)-s\big)\right]\,,
\label{eq:PDF}
}
or, equivalently, through its Laplace transform,
\beeq{
\Phi_{\omega,T}(\lambda)=\mathbb{E}_{\{X(t)\}}\left[e^{-\lambda S(\omega,T)}\right]\,,
\label{eq:MGFmt}
}
from which the moments of $S(\omega,T)$ follow by differentiation,
\beeq{
\mathbb{E}_{\{X(t)\}}\left[S^n(\omega,T)\right]=(-1)^n\frac{\partial^n}{\partial \lambda^n}\Phi_{\omega,T}(\lambda)\Big|_{\lambda=0}\,.
}
The Laplace transform \eqref{eq:MGFmt} will be the basic generating object in the single-frequency analysis of Section~\ref{onefrequency}.

More generally, spectral information is usually extracted at several frequencies. Given a set $\bm{\omega}\equiv(\omega_1,\ldots,\omega_L)$, we consider the collection of random variables $\{S(\omega_i,T)\}_{i=1}^{L}$. Their joint statistics are encoded in the corresponding joint PDF,
\beeq{
\rho_{\bm{\omega},T}(\bm{s})\equiv\mathbb{E}_{\{X(t)\}}\left[\prod_{i=1}^{L}\delta\!\big(S(\omega_i,T)-s_i\big)\right]\,,
\label{eq:jPDF}
}
or, equivalently, in the multivariate Laplace transform
\beeq{
\Phi_{\bm{\omega},T}(\bm{\lambda})=\mathbb{E}_{\{X(t)\}}\left[\exp\left( -\sum_{i=1}^{L}\lambda_i S(\omega_i,T)\right)\right]\,,
\label{eq:jMGF}
}
where we have defined $\bm{s}=(s_1,\ldots,s_L)$ and $\bm{\lambda}=(\lambda_1,\ldots,\lambda_L)$. As in the single-frequency case, mixed moments are obtained by differentiating \eqref{eq:jMGF},
\beeq{
\mathbb{E}_{\{X(t)\}}\left[\prod_{i=1}^L S(\omega_i,T)^{n_i}\right]=(-1)^{\sum_i n_i}\left.\frac{\partial^{\sum_i n_i}\Phi_{\bm\omega,T}(\bm\lambda)}{\partial \lambda_1^{n_1}\cdots \partial \lambda_L^{n_L}}\right|_{\bm\lambda=\bm 0}\,.
\label{eq:mixed_moments}
}

To obtain explicit finite-$T$ expressions, we now specialize to the Ornstein--Uhlenbeck (OU) process, motivated in particular by the dynamics of an overdamped Brownian particle in a harmonic trap. The process $X(t)$ obeys
\beeq{
\frac{d}{dt}X(t)=-\frac{1}{\tau}X(t)+\sqrt{2D} W_x(t)\,,
\label{eq:OU}
}
where $W_x(t)$ denotes Gaussian white noise with zero mean and $\bracket{ W_x(t)W_x(t')}=\delta(t-t')$. Here $\tau$ is the relaxation time and $D$ the diffusion constant. A formal solution of \eqref{eq:OU} is
\beeq{
X(t)=X_0 e^{-t/\tau} +\sqrt{2D}\int_0^t ds e^{-(t-s)/\tau}W_x(s)\,,
\label{eq:solOU}
}
with $X_0=X(0)$.

The finite-$T$ distributions introduced above are well defined for both stationary and nonstationary Gaussian preparations. If $X_0$ is Gaussian, independent of the driving noise, with mean $m_0$ and variance $V_0$, then $\mathbb{E}_{\{X(t)\}}[X(t)]=m_0 e^{-t/\tau}$ and the two-time covariance
\beeq{
C(t,t')\equiv \mathrm{Cov}\big(X(t),X(t')\big)= D\tau e^{-|t-t'|/\tau} + (V_0-D\tau) e^{-(t+t')/\tau}
}
contains the dependence on the initial preparation only through a transient term that decays on the relaxation scale $\tau$ (see Appendix~\ref{appA} for details). In particular, the asymptotic limit $\mu(\omega)=\lim_{T\to\infty}\mathbb{E}[S(\omega,T)]$ is independent of the initial preparation, whereas finite-$T$ corrections need not be. In the main text, unless stated otherwise, we work with centered deterministic initialization, $m_0=0$ and $X_0=0$, in order to keep the formulas compact. Note that the stationary Gaussian initialization is recovered by choosing $m_0=0$ and $V_0=D\tau$.

Finally, it is useful to make explicit the bridge between the continuous-time finite-window objects defined here and the sampled-data inference problem considered later. Sections~\ref{onefrequency} and \ref{multiplefrequency} analyze $S(\omega,T)$ for arbitrary $\omega$ in continuous time, since this is the natural setting in which the exact finite-$T$ multispectral structure can be derived. Section~\ref{sec:inference}, by contrast, works with a uniformly sampled record $x_n\equiv X(n\Delta t)$, $n=0,\ldots,N-1$, with $T=N\Delta t$, and with the corresponding discrete Fourier frequencies $\omega_k=2\pi k/T$. For the OU process, we recall that the sampled sequence is exactly an AR(1) process, so the likelihoods studied in Section~\ref{sec:inference} should be viewed as sampled-data realizations of the same finite-window spectral framework introduced here.

%%%%%%%%%%%%%%%%%%%%%%%%%%%%%%%%%%%%%%%%%%%%%%%%%%%
\section{Single-frequency analysis}
\label{onefrequency}
%%%%%%%%%%%%%%%%%%%%%%%%%%%%%%%%%%%%%%%%%%%%%%%%%%%
To derive the moment generating function (MGF) and, from it, the PDF of the finite-time spectral estimator $S(\omega,T)$, we exploit the Gaussian structure of the underlying process. For the centered deterministic initialization that we adopt here for simplicity, a convenient starting point is to rewrite $S(\omega,T)$ as the sum of squares of two linear functionals of the trajectory,
\beeq{
S(\omega,T) =\frac{1}{T}\left(\int_0^T dt \cos(\omega t)X(t)\right)^2+\frac{1}{T}\left(\int_0^T dt \sin(\omega t)X(t)\right)^2 \,.
}
This representation shows that the single-frequency problem is governed by the cosine and sine Fourier projections of the trajectory on the observation window. Since the OU process is Gaussian, these two projections form a centered bivariate Gaussian vector, so the full law of $S(\omega,T)$ is determined by a $2\times 2$ covariance structure, in agreement with the general centered-Gaussian framework of single-trajectory spectral fluctuations discussed in \cite{squarcini2022noise}.

The quadratic form above allows us to linearize the exponential appearing in the MGF by means of a Hubbard--Stratonovich transformation, yielding
\beeq{
\Phi_{\omega,T}(\lambda)=\int G(v,w)\mathbb{E}_{\{X(t)\}}\left[\exp\left(i\sqrt{\frac{2\lambda}{T}}\int_0^T dt [v\cos(\omega t)+w\sin(\omega t)]X(t)\right)\right]\,,
\label{eq:a}
}
where $G(v,w)=\exp\left[-(v^2+w^2)/2\right]dvdw/(2\pi)$ denotes the standard Gaussian measure over the auxiliary variables $(v,w)$. Inserting the formal solution \eqref{eq:solOU} into \eqref{eq:a} and performing the Gaussian average over the noise history, one arrives at three scalar coefficients,
\beeq{
A^{(vv)}_{\omega,T}&=\int_0^T ds \left(\int_s^T dt \cos(\omega t)e^{-(t-s)/\tau}\right)^2\,,\\
A^{(ww)}_{\omega,T}&=\int_0^T ds \left(\int_s^T dt \sin(\omega t)e^{-(t-s)/\tau}\right)^2\,,\\
A^{(vw)}_{\omega,T}&=\int_0^T ds \left(\int_s^T dt \cos(\omega t)e^{-(t-s)/\tau}\right)\left(\int_s^T dt' \sin(\omega t')e^{-(t'-s)/\tau}\right)\,.
}
These integrals can be explicitly carried out, but the resulting closed forms are unwieldy so we collect them in Appendix~\ref{appA}. It is convenient, however,  to assemble them into the symmetric matrix
\beeq{
\bm{A}_{\omega,T}=\left(
\begin{array}{cc}
A^{(vv)}_{\omega,T} & A^{(vw)}_{\omega,T}\\
A^{(vw)}_{\omega,T} & A^{(ww)}_{\omega,T}
\end{array}
\right)\,,
}
so that the MGF takes the compact form
\beeq{
\Phi_{\omega,T}(\lambda)=\left[\det\left(\mathbb{I}+\lambda\frac{4D}{T}\bm{A}_{\omega,T}\right)\right]^{-1/2}=\left[1+\lambda\frac{4D}{T} \left(A^{(vv)}_{\omega,T}+A^{(ww)}_{\omega,T}\right)+\lambda^2\left(\frac{4D}{T}\right)^2\left(A^{(vv)}_{\omega,T}A^{(ww)}_{\omega,T}-[A^{(vw)}_{\omega,T}]^2\right)\right]^{-1/2}\,.
\label{eq:MGFfinal}
}
This makes explicit that the single-frequency distribution is controlled entirely by the two-dimensional Gaussian covariance structure encoded in $\bm{A}_{\omega,T}$. In particular,  the mean value and the variance are  most naturally written as the trace invariants
\beeq{
\mu_{\omega,T}&\equiv\mathbb{E}_{\{X(t)\}}[S(\omega,T)]=\frac{2D}{T}\mathrm{Tr}\,\bm{A}_{\omega,T}=\frac{2D}{T} \left(A^{(vv)}_{\omega,T}+A^{(ww)}_{\omega,T}\right)\,,\\
\sigma^2_{\omega,T}&\equiv \text{Var}[S(\omega,T)]=\frac{1}{2}\left(\frac{4D}{T}\right)^2\mathrm{Tr}\,\bm{A}_{\omega,T}^2=\frac{1}{2}\left(\frac{4D}{T}\right)^2\left\{ [A^{(vv)}_{\omega,T}]^2+[A^{(ww)}_{\omega,T}]^2+2[A^{(vw)}_{\omega,T}]^2\right\}\,,
}
respectively.

To quantify the magnitude of the fluctuations of $S(\omega,T)$ relative to its mean value, we introduce the coefficient of variation $\gamma_{\omega,T}\equiv \sigma_{\omega,T}/\mu_{\omega,T}$ which, in terms of $\bm{A}_{\omega,T}$, reads
\beeq{
\gamma^2_{\omega,T}=2\frac{\mathrm{Tr}\,\bm{A}_{\omega,T}^2}{\left(\mathrm{Tr}\,\bm{A}_{\omega,T}\right)^2}\,.
}
Since $\bm{A}_{\omega,T}$ is positive semidefinite, its two eigenvalues are nonnegative, and it follows immediately that
$1\leq \gamma_{\omega,T}\leq \sqrt{2}$ \footnote{The lower bound results when the two eigenvalues are equal while the upper bound is obtained from the fact that $x^2+y^2\leq (x+y)^2$ for $x$ and $y$ positive.}. These bounds are remarkable  since they indicate that, even at a fixed frequency, single-trajectory spectral estimators remain broadly distributed at finite $T$, a feature that is consistent with the general noise-to-signal picture for centered Gaussian processes \cite{squarcini2022noise}.

\begin{figure}
\begin{center}
\includegraphics[width=\linewidth]{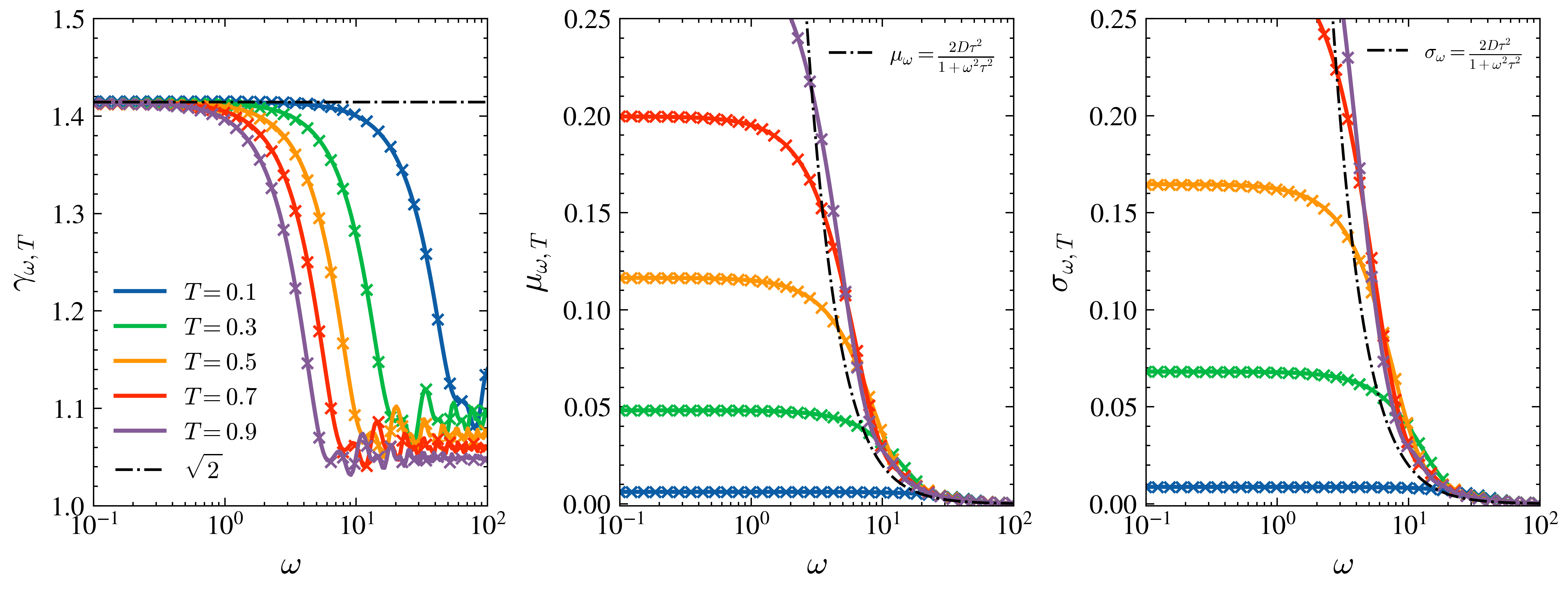}
\caption{Plot of the coefficient of variation $\gamma_{\omega,T}$, mean value $\mu_{\omega,T}$ and standard deviation $\sigma_{\omega,T}$ of $S(\omega,T)$ as a function of the angular frequency $\omega$ for various acquisition times $T=0.1$, $0.3$, $0.5$, $0.7$, and $0.9$. For these plots the diffusion coefficient and relaxation time are fixed to $D=1$ and $\tau=1$. Markers correspond to Monte Carlo simulations obtained by discretizing the stochastic differential equation \eqref{eq:OU} with time step $\Delta t=10^{-3}$ and averaging over $10^5$ trajectories.}
\label{fig1}
\end{center}
\end{figure}

Figure~\ref{fig1} compares these predictions with Monte Carlo simulations obtained by applying the Euler--Maruyama scheme to the stochastic differential equation \eqref{eq:OU} with time step $\Delta t=10^{-3}$ and averaging over $10^5$ trajectories. The agreement confirms the finite-$T$ expressions above over the full frequency range shown.

Two limiting cases provide additional checks. In the limit of an infinitely large relaxation time, $\tau\to\infty$, the OU process reduces to the free Brownian motion, and the expressions for the mean value and variance of $S(\omega,T)$ become
\beeq{
\lim_{\tau\to\infty} \mu_{\omega,T}&=\frac{4D}{\omega^2}\left(1-\frac{\sin(\omega T)}{\omega T}\right)\,,\\
\lim_{\tau\to\infty} \sigma^2_{\omega,T}&=\frac{2D^2}{\omega^6 T^2}\Bigg(17+10 \omega^2 T^2-16\cos(\omega T)-\cos(2\omega T)+2\omega T[\sin(2\omega T)-12\sin(\omega T)]\Bigg)\,,\\
}
in agreement with the results reported in Ref.~\cite{krapf2018power}. Moreover, in the limit of infinite acquisition time $T\to\infty$ we obtain the following simple expressions for the first two cumulants:
\beeq{
\mu_{\omega}\equiv\lim_{T\to\infty}\mu_{\omega,T}=2D\frac{\tau^2}{1+\omega^2\tau^2}\,,\quad\quad
\sigma^2_{\omega}\equiv\lim_{T\to\infty}\sigma^2_{\omega,T}=\left(2D\frac{\tau^2}{1+\omega^2\tau^2}\right)^2\,.
}
Thus the coefficient of variation tends to unity as $T\to\infty$, consistently with an asymptotic exponential single-frequency law, that we derive below.

It is useful to rewrite the MGF directly in terms of the first two cumulants,
\beeq{
\Phi_{\omega,T}(\lambda)=\left[1+2\lambda\mu_{\omega,T}+\lambda^2\left(2\mu_{\omega,T}^2-\sigma^2_{\omega,T}\right)\right]^{-1/2}
=\left[1+2\lambda\mu_{\omega,T}+\lambda^2\mu_{\omega,T}^2\left(2-\gamma^2_{\omega,T}\right)\right]^{-1/2}\,.
}
In this form the inverse Laplace transform can be performed explicitly, yielding
\beeq{
\rho_{\omega,T}(s)&=\frac{1}{\mu_{\omega,T}\sqrt{2-\gamma^2_{\omega,T}}}e^{-\frac{s}{\mu_{\omega,T}(2-\gamma^2_{\omega,T})}}I_0\left(s\frac{\sqrt{\gamma^2_{\omega,T}-1}}{\mu_{\omega,T}(2-\gamma^2_{\omega,T})}\right)\,,
\label{eq:pdffinal}
}
where $I_\nu(x)$ is the modified Bessel function of the first kind. Equation \eqref{eq:pdffinal} has the same Bessel-type structure found for other single-trajectory spectral observables, for instance in the case of scaled Brownian motion \cite{sposini2019single}, and is consistent with the broader centered-Gaussian framework discussed in \cite{squarcini2022noise}. For the OU process, all the dependence on $\omega$, $T$, $D$, and $\tau$ enters through $\mu_{\omega,T}$ and $\gamma_{\omega,T}$.

\begin{figure}
\begin{center}
\includegraphics[width=\linewidth]{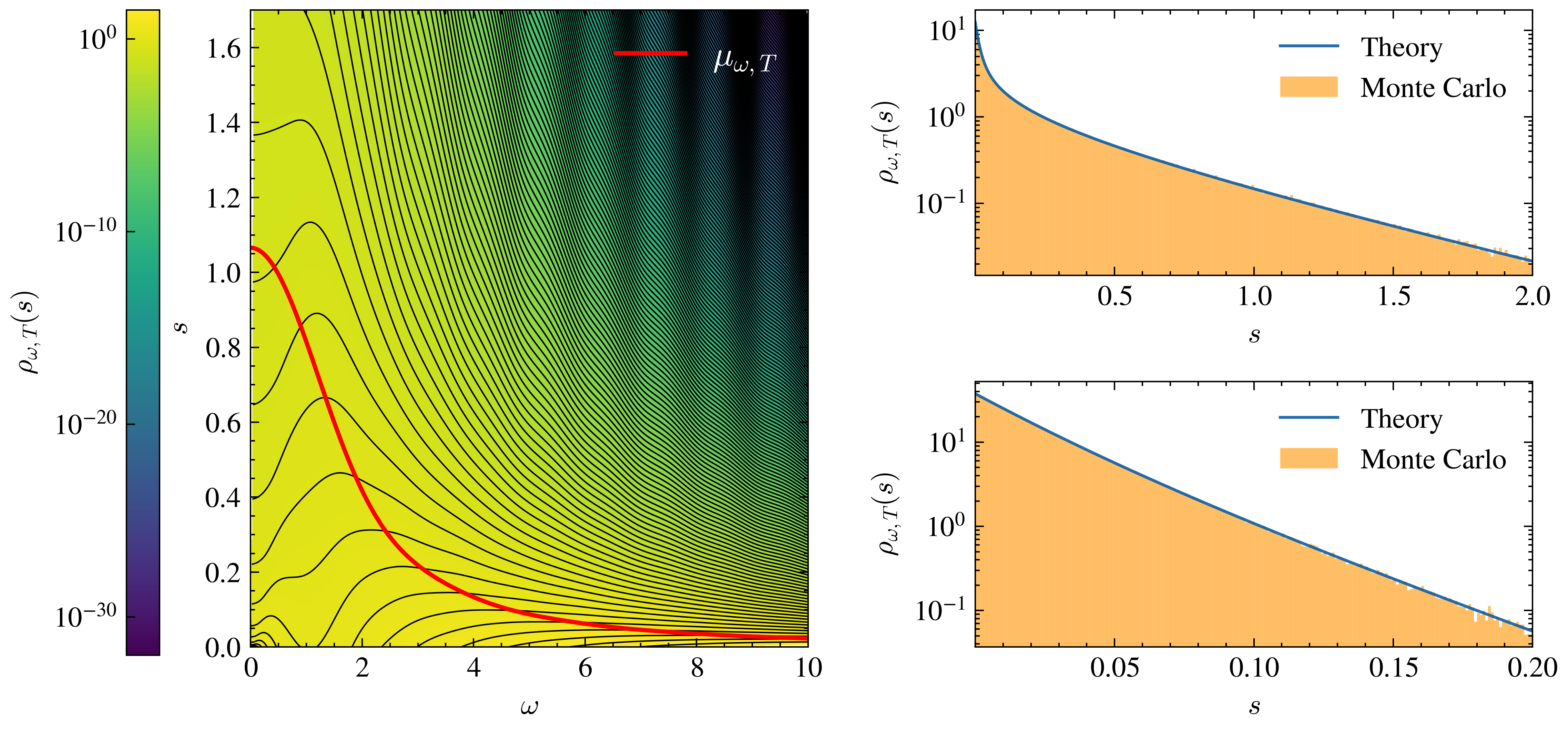}
\end{center}
\caption{Left panel: density plot of $\rho_{\omega,T}(s)$ in the $(\omega,s)$-plane for $T=3$, $\tau=1$, and $D=1$. Right panel: comparison between Monte Carlo simulations and the theoretical prediction for $\rho_{\omega,T}(s)$ at $\omega=1$ (top) and $\omega=10$ (bottom), with $T=1$, $\tau=1$, and $D=1$.}
\label{fig2}
\end{figure}

In the limit of infinite acquisition time $T\to\infty$ the expression simplifies to an exponential law
\beeq{
\rho_\omega(s)\equiv \lim_{T\to\infty} \rho_{\omega,T}(s)=\frac{1}{\mu_{\omega}}e^{-\frac{s}{\mu_{\omega}}}\,,
}
where we used that the coefficient of variation tends to unity in this limit and $I_0(0)=1$. In Fig.~\ref{fig2} we show representative plots of $\rho_{\omega,T}(s)$.

%%%%%%%%%%%%%%%%%%%%%%%%%%%%%%%%%%%%%%%%%%%%%%%%%%%
\section{Multiple-frequency analysis}
\label{multiplefrequency}
%%%%%%%%%%%%%%%%%%%%%%%%%%%%%%%%%%%%%%%%%%%%%%%%%%%
At finite observation time, the relevant spectral object is not a collection of independent ordinates but the full multispectral structure induced by windowing. Our main result here is that, for the OU process, this structure admits an exact finite-$T$ covariance-explicit formulation. This provides a natural bridge from the single-frequency laws derived in Section~\ref{onefrequency} to correlated frequency-domain statistics, from where one can  then introduce likelihoods useful for inferential processes.

To keep the underlying Gaussian structure manifest, we begin with the linear Fourier projections of the trajectory. For each frequency $\omega_i$ we define
\beeq{
Z_{c,i}\equiv \int_0^Tdt\cos(\omega_i t) X(t)\,,\qquad Z_{s,i}\equiv \int_0^T dt \sin(\omega_i t)X(t)\,,\qquad i=1,\dots,L\,.
\label{eq:definition_Z}
}
We then collect them into a single $2L$-dimensional real vector
\beeq{
Z\equiv (Z_{c,1},\dots,Z_{c,L},Z_{s,1},\dots,Z_{s,L})^{T}\,.
}
By construction, the finite-$T$ spectral estimators at the selected frequencies are the quadratic forms
\beeq{
S(\omega_i,T)=\frac{1}{T}\left(Z_{c,i}^2+Z_{s,i}^2\right)\,,\qquad i=1,\dots,L\,.
\label{eq:S_from_Z}
}
Thus the collection $\{S(\omega_i,T)\}_{i=1}^L$ may be viewed as the set of squared radii of the $L$ two-dimensional vectors $(Z_{c,i},Z_{s,i})$, rescaled by $1/T$.

For the centered deterministic initialization used here, the $2L$-dimensional real vector $Z$ is exactly multivariate normal at finite $T$:
\beeq{
Z\sim \mathcal N(0,\Sigma),\qquad \Sigma\equiv \mathbb{E}_{\{X(t)\}}\{ZZ^{T}\}\,.
\label{eq:Z_gaussian}
}
One can show (see Appendix~\ref{appD}) that the entries of $\Sigma$ are expressed in terms of the kernels
\beeq{
A^{(v_iv_j)}_{\omega_i,\omega_j,T}&\equiv\int_0^T ds\left(\int_s^T dt \cos(\omega_i t)e^{-(t-s)/\tau}\right)\left(\int_s^T dt \cos(\omega_j t)e^{-(t-s)/\tau}\right)\,,\\
A^{(w_iw_j)}_{\omega_i,\omega_j,T}&\equiv\int_0^T ds\left(\int_s^T dt \sin(\omega_i t)e^{-(t-s)/\tau}\right)\left(\int_s^T dt \sin(\omega_j t)e^{-(t-s)/\tau}\right)\,,\\
A^{(v_iw_j)}_{\omega_i,\omega_j,T}&\equiv\int_0^T ds\left(\int_s^T dt \cos(\omega_i t)e^{-(t-s)/\tau}\right)\left(\int_s^T dt \sin(\omega_j t)e^{-(t-s)/\tau}\right)\,,
\label{eq:A_multi_frequency}
}
with $A^{(w_iv_j)}_{\omega_i,\omega_j,T}=A^{(v_j w_i)}_{\omega_j,\omega_i,T}$. More precisely,
\beeq{
\mathbb{E}_{\{X(t)\}}\{Z_{c,i}Z_{c,j}\}=2D A^{(v_iv_j)}_{\omega_i,\omega_j,T}\,,\qquad  \mathbb{E}_{\{X(t)\}}\{Z_{s,i}Z_{s,j}\}=2D A^{(w_iw_j)}_{\omega_i,\omega_j,T}\,,\qquad 
\mathbb{E}_{\{X(t)\}}\{Z_{c,i}Z_{s,j}\}=2DA^{(v_iw_j)}_{\omega_i,\omega_j,T}\,,
\label{eq:Sigma_entries}
}
with $\mathbb{E}_{\{X(t)\}}\{Z_{s,i}Z_{c,j}\}=2DA^{(w_iv_j)}_{\omega_i,\omega_j,T}$. Writing these entries in block form, one obtains
\beeq{
\mathcal{A}=\begin{pmatrix}
    \bm{A}^{(vv)}&\bm{A}^{(vw)}\\
    \left[\bm{A}^{(vw)}\right]^T&\bm{A}^{(ww)}
\end{pmatrix}\,,
}
and therefore
\beeq{
\Sigma = 2D\mathcal A\,.
\label{eq:Sigma_equals_A}
}
This identity is the basic finite-$T$ multispectral result: the full joint structure is encoded in the covariance matrix $\Sigma$, and the asymptotic independence picture emerges only when its off-diagonal frequency blocks become negligible.

The Gaussian law \eqref{eq:Z_gaussian} yields an explicit correlated likelihood for the  data encoded in the $Z$ projections:
\beeq{
p_{\bm{\omega},T}(Z\mid D,\tau)=\frac{1}{(2\pi)^{L}\sqrt{\det\Sigma}}\exp\left(-\frac{1}{2}Z^{T}\Sigma^{-1}Z\right)=\frac{1}{(2\pi)^{L}\sqrt{\det(2D\mathcal A)}}\exp\left(-\frac{1}{4D}Z^{T}\mathcal A^{-1}Z\right)\,.
\label{eq:liftedGaussian_multi}
}
This representation is particularly convenient for inference because it keeps the multispectral correlations explicit through $\Sigma$ (or equivalently $\mathcal A$) while remaining fully regular. Moreover, this same representation provides a compact expression for the multispectral Laplace transform of the quadratic observables \eqref{eq:S_from_Z}. Let
\beeq{
\Lambda\equiv \mathrm{diag}(\lambda_1,\dots,\lambda_L,\lambda_1,\dots,\lambda_L)\,,
}
so that $\sum_{i=1}^L\lambda_i S(\omega_i,T)=(1/T)\,Z^{T}\Lambda Z$. Then
\beeq{
\Phi_{\bm{\omega},T}(\bm{\lambda}) =\mathbb{E}_{\{X(t)\}}\left[\exp\left(-\sum_{i=1}^L\lambda_i S(\omega_i,T)\right)\right]=\det\left(I+\frac{2}{T}\Sigma\Lambda\right)^{-1/2}
=\det\left(I+\frac{4D}{T}\mathcal A\Lambda\right)^{-1/2}\,.
\label{eq:Laplace_det_Z}
}
Equations \eqref{eq:liftedGaussian_multi}  and \eqref{eq:Laplace_det_Z} are the exact finite-$T$ multispectral results of this work, which will be further used throughout the remainder of the paper. These  results go beyond single-frequency laws and pairwise correlation diagnostics: they yield a covariance-explicit joint-law / likelihood framework for the entire multispectral collection.

Although no scalar coefficient of variation is canonically associated with the full vector $\{S(\omega_i,T)\}_{i=1}^{L}$, Eq.~\eqref{eq:Laplace_det_Z} immediately yields one for the total spectral power over the selected frequencies,
\beeq{
S_{\rm tot}(\bm\omega,T)\equiv \sum_{i=1}^{L}S(\omega_i,T)=\frac{1}{T}Z^{T}Z\,.
\label{eq:Stot_definition}
}
Indeed, setting $\lambda_1=\cdots=\lambda_L=\lambda$ in Eq.~\eqref{eq:Laplace_det_Z} gives
\beeq{
\Phi^{\rm tot}_{\bm\omega,T}(\lambda)\equiv \mathbb{E}_{\{X(t)\}}\left[e^{-\lambda S_{\rm tot}(\bm\omega,T)}\right]
=\det\left(I+\frac{4D\lambda}{T}\mathcal A\right)^{-1/2}\,,
\label{eq:Stot_laplace}
}
and therefore
\beeq{
\mu^{\rm tot}_{\bm\omega,T}=\frac{2D}{T}\mathrm{Tr}\mathcal{A}\,,\qquad
(\sigma^{\rm tot}_{\bm\omega,T})^2=\frac{8D^2}{T^2}\mathrm{Tr}\mathcal{A}^2\,,\qquad
(\gamma^{\rm tot}_{\bm\omega,T})^2\equiv \frac{(\sigma^{\rm tot}_{\bm\omega,T})^2}{(\mu^{\rm tot}_{\bm\omega,T})^2}
=2\frac{\mathrm{Tr}\mathcal{A}^2}{\left(\mathrm{Tr}\mathcal A\right)^2}\,.
\label{eq:Stot_CV}
}
Since $\mathcal{A}$ is positive semidefinite, it follows that
\beeq{
\frac{1}{\sqrt{L}}\leq \gamma^{\rm tot}_{\bm\omega,T}\leq \sqrt{2}\,.
\label{eq:Stot_CV_bounds}
}

Thus, even after aggregation over $L$ frequencies, the total spectral power remains broadly distributed at finite $T$; summation over frequencies can at best reduce the relative fluctuations as $L^{-1/2}$. Appendix~\ref{appF} extends this trace formula to weighted multispectral powers $Y_{\bm q}(\bm\omega,T)=\sum_{i=1}^{L}q_i S(\omega_i,T)$ with $q_i\geq 0$, showing that their coefficient of variation is governed by the effective rank of the corresponding weighted multispectral matrix and, in particular, satisfies $1/m_{\bm q}\leq \gamma_{\bm q,\bm\omega,T}^{2}\leq 2$, where $m_{\bm q}\equiv \#\{i:q_i>0\}$. The same appendix also proves the comparison $\Phi_{\bm\omega,T}(\bm\lambda)\geq \prod_{i=1}^{L}\Phi_{\omega_i,T}(\lambda_i)$ with the fully factorized exact single-frequency approximation.

\begin{figure}[H]
\begin{center}
\includegraphics[width=\linewidth]{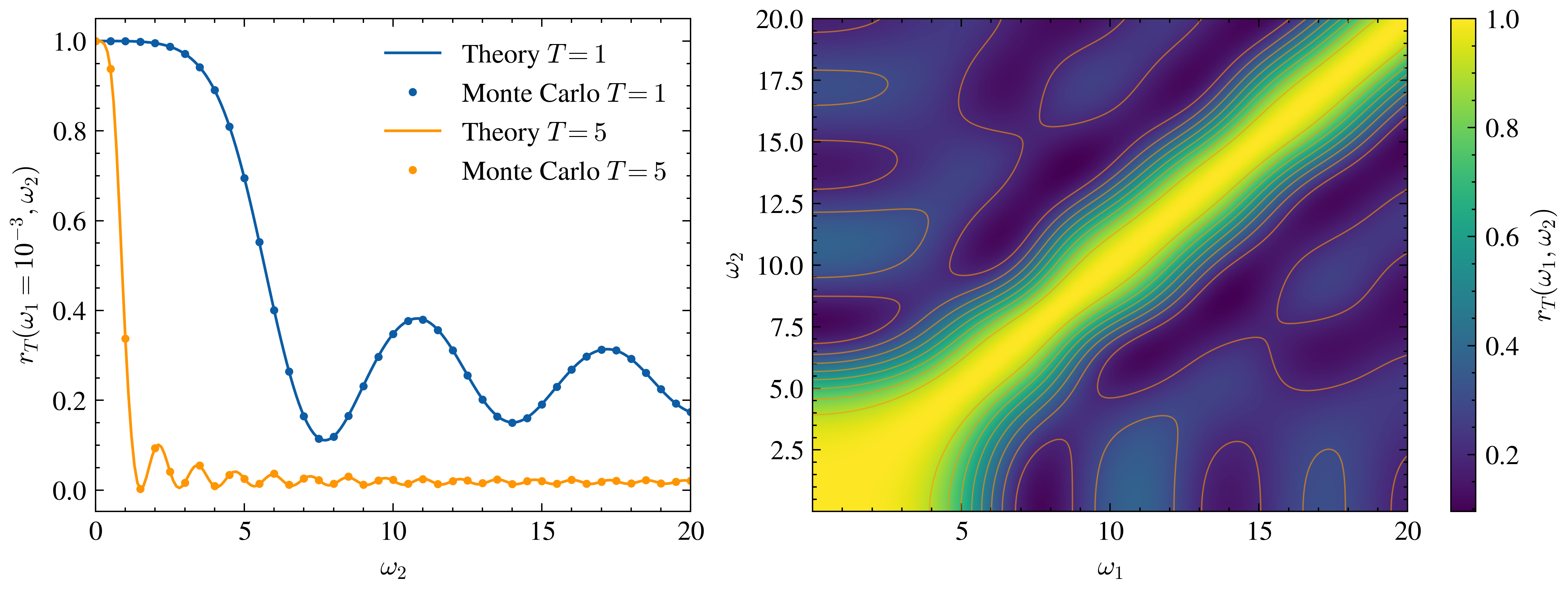}
\end{center}
\caption{Correlation coefficient $r_T(\omega_1,\omega_2)$ between $S(\omega_1,T)$ and $S(\omega_2,T)$. Monte Carlo (MC) data are obtained from Euler-discretized OU trajectories with $\Delta t=10^{-3}$ using $10^{6}$ independent realizations, and $r_T$ is estimated from the resulting samples of $\{S(\omega_i,T)\}$. Parameters are fixed to $D=\tau=1$. Right panel: theoretical prediction for $r_T(\omega_1,\omega_2)$ at $T=1$ in the $(\omega_1,\omega_2)$ plane.}
\label{fig3}
\end{figure}

Differentiating Eq.~\eqref{eq:Laplace_det_Z}, or equivalently applying Wick's theorem to the Gaussian vector $Z$, gives the lowest multispectral moments. In particular,
\beeq{
\mathbb{E}_{\{X(t)\}}\left\{S(\omega_i,T)\right\}&=\frac{2D}{T}\Big(A^{(v_iv_i)}_{\omega_i,\omega_i,T}+A^{(w_iw_i)}_{\omega_i,\omega_i,T}\Big)\,,
}
and
\beeq{
\mathbb{E}_{\{X(t)\}}\left\{[S(\omega_i,T)]^2\right\}&=\left(\frac{2D}{T}\right)^2\Bigg\langle\Big(A^{(v_iv_i)}_{\omega_i,\omega_i,T} v_i^2+A^{(w_iw_i)}_{\omega_i,\omega_i,T} w_i^2+2A^{(v_iw_i)}_{\omega_i,\omega_i,T} v_i w_i\Big)^2\Bigg\rangle_{\bm{v},\bm{w}}\\
&=\left(\frac{2D}{T}\right)^2\Big\{3\big[A^{(v_iv_i)}_{\omega_i,\omega_i,T}\big]^2+3\big[A^{(w_iw_i)}_{\omega_i,\omega_i,T}\big]^2+4\big[A^{(v_iw_i)}_{\omega_i,\omega_i,T}\big]^2+2A^{(v_iv_i)}_{\omega_i,\omega_i,T} A^{(w_iw_i)}_{\omega_i,\omega_i,T}\Big\}\,,
}
while, for $\omega_i\neq \omega_j$,
\beeq{
\mathbb{E}_{\{X(t)\}}\left\{S(\omega_i,T)\,S(\omega_j,T)\right\}&=\left(\frac{2D}{T}\right)^2
\Big(A^{(v_iv_i)}_{\omega_i,\omega_i,T}+A^{(w_iw_i)}_{\omega_i,\omega_i,T}\Big)
\Big(A^{(v_jv_j)}_{\omega_j,\omega_j,T}+A^{(w_jw_j)}_{\omega_j,\omega_j,T}\Big)\\
&+2\left(\frac{2D}{T}\right)^2\Big(\big[A^{(v_iv_j)}_{\omega_i,\omega_j,T}\big]^2+\big[A^{(w_iw_j)}_{\omega_i,\omega_j,T}\big]^2+\big[A^{(v_iw_j)}_{\omega_i,\omega_j,T}\big]^2+\big[A^{(w_iv_j)}_{\omega_i,\omega_j,T}\big]^2\Big)\,,\qquad i\neq j\,.
}
It is then convenient to introduce the covariance function
\beeq{
C_T(\omega_i,\omega_j)&\equiv \mathbb{E}_{\{X(t)\}}\left\{S(\omega_i,T)S(\omega_j,T)\right\}
-\mathbb{E}_{\{X(t)\}}\left\{S(\omega_i,T)\right\}\mathbb{E}_{\{X(t)\}}\left\{S(\omega_j,T)\right\}\\
&=2\left(\frac{2D}{T}\right)^2\Big(\big[A^{(v_iv_j)}_{\omega_i,\omega_j,T}\big]^2+\big[A^{(w_iw_j)}_{\omega_i,\omega_j,T}\big]^2+\big[A^{(v_iw_j)}_{\omega_i,\omega_j,T}\big]^2+\big[A^{(w_iv_j)}_{\omega_i,\omega_j,T}\big]^2\Big)\,.
}
Figure~\ref{fig3} compares this prediction, displayed through $r_T(\omega_i,\omega_j)\equiv C_T(\omega_i,\omega_j)/(\sigma_{\omega_i,T}\sigma_{\omega_j,T})$, the correlation coefficient, with Monte Carlo data. The agreement shows that finite-time spectra evaluated at different frequencies are generally correlated and that these correlations vanish only asymptotically.

These correlations originate from the finite observation window. Indeed, since $S(\omega,T)=\big|\int_0^T dt e^{i\omega t}X(t)\big|^2/T$, restricting the trajectory to $[0,T]$ amounts to multiplying by the rectangular time window $w_T(t)=\mathbf 1_{[0,T]}(t)$. In the frequency domain this becomes the convolution with $W_T(\Omega)=\int_0^T dt\,e^{i\Omega t} = e^{i\Omega T/2}\,\frac{2\sin(\Omega T/2)}{\Omega}$, so the finite-time Fourier projections mix spectral content over a scale $\Delta\omega\sim 1/T$. For the OU process considered here, this produces inter-frequency correlations that are strongest for $|\omega_i-\omega_j|=O(1/T)$ and decay beyond the leakage scale.  On the natural Fourier grid $\omega_k=2\pi k/T$, neighboring bins are therefore separated precisely on the scale on which the covariance is most significant, which explains why the multispectral covariance is naturally close to block-banded in the ordered frequency index. Appendix~\ref{appE} explains why this localization does not amount to a universal window-only decorrelation in the physical Fourier basis.

This observation suggests a controlled hierarchy of diagonal-in-frequency approximations obtained by sparsifying $\Sigma$ at the level of its $2\times 2$ frequency blocks. Ordering the frequencies as $\omega_1<\cdots<\omega_L$ and partitioning them into contiguous groups $\mathcal B_1,\mathcal B_2,\ldots$ of size $m$ (with the last group possibly smaller), we define $\Sigma^{(m)}$ as the block-diagonal matrix obtained by retaining the full $2|\mathcal B_\alpha|\times 2|\mathcal B_\alpha|$ principal submatrix of $\Sigma$ within each group $\mathcal B_\alpha$ and setting to zero all blocks that couple distinct groups. This construction is monotone in $m$: the limit $m=1$ yields complete factorization across frequencies, while $m=L$ recovers the exact covariance $\Sigma$.

Within the $Z$-projection representation, $\Sigma^{(m)}$ induces the explicit block-factorized Gaussian likelihood
\beeq{
p^{(m)}_{\bm\omega,T}(\bm z\mid D,\tau)
\equiv \mathcal N\!\left(\bm z;0,\Sigma^{(m)}(D,\tau)\right)
=\prod_{\alpha}
\frac{1}{(2\pi)^{|\mathcal B_\alpha|}\sqrt{\det \Sigma_{\mathcal B_\alpha}(D,\tau)}}
\exp\!\left[-\frac{1}{2}\bm z_{\mathcal B_\alpha}^{T}\Sigma_{\mathcal B_\alpha}(D,\tau)^{-1}\bm z_{\mathcal B_\alpha}\right],
\label{eq:block_lifted_like}
}
where $\bm z_{\mathcal B_\alpha}$ collects the $(Z_{c,i},Z_{s,i})$ variables for $i\in\mathcal B_\alpha$, and $\Sigma_{\mathcal B_\alpha}$ is the corresponding principal submatrix of $\Sigma$. Since $S(\omega_i,T)=(Z_{c,i}^2+Z_{s,i}^2)/T$, the same hierarchy induces a family of approximate joint densities for $\bm S=(S(\omega_1,T),\ldots,S(\omega_L,T))$ in which independence holds across frequency groups:
\beeq{
\rho^{(m)}_{\bm\omega,T}(\bm s\mid D,\tau)=\prod_{\alpha}\rho_{\mathcal{B}_\alpha,T}\left(\bm s_{\mathcal B_\alpha}\mid D,\tau\right)\,,
\label{eq:block_rho_S}
}
with $\rho_{\mathcal B_\alpha,T}$ denoting the exact joint density restricted to the subset $\mathcal B_\alpha$. In particular, the fully factorized limit is
\beeq{
\rho^{(1)}_{\bm\omega,T}(\bm s \mid D,\tau)=\prod_{i=1}^L \rho_{\omega_i,T}(s_i\mid D,\tau)\,,
\label{eq:full_factor_S}
}
namely the product of the exact finite-$T$ single-frequency laws given by Eq. \eqref{eq:pdffinal}, which we derived in Section~\ref{onefrequency}. Thus $m=1$ is already more informative than the classical Whittle approximation as it factorizes across frequencies, but it retains the exact finite-$T$ single-frequency law at each frequency.

Indeed, recall that the classical Whittle construction appears only after a further large-$T$ simplification. On a discrete Fourier grid $\omega_k=2\pi k/T$ (or $\omega_k=2\pi k/(N\Delta t)$ in discrete time with $T=N\Delta t$), the standard working assumption is that Fourier coefficients, and hence periodogram ordinates, are effectively independent across $k$. In the present language this corresponds to the diagonal-in-frequency endpoint together with the asymptotic exponential approximation for each ordinate, that is,
\beeq{
p\left(I(\omega_k)\mid D,\tau\right)\simeq \frac{1}{S_{D,\tau}(\omega_k)}\exp\left[-\frac{I(\omega_k)}{S_{D,\tau}(\omega_k)}\right]\,,\qquad I(\omega_k)\equiv \frac{1}{T}\Big|\int_0^T dt\,e^{i\omega_k t}X(t)\Big|^2\,,
\label{eq:periodogram_exp}
}
which yields the associated factorized Whittle likelihood
\beeq{
\mathcal{L}_{\rm W}(D,\tau) \simeq \prod_{k\in\mathcal K}\frac{1}{S_{D,\tau}(\omega_k)}\exp\left[-\frac{I(\omega_k)}{S_{D,\tau}(\omega_k)}\right]\,.
\label{eq:Whittle_like}
}
From the finite-$T$ perspective developed here, Eqs.~\eqref{eq:periodogram_exp} and \eqref{eq:Whittle_like} arise only after neglecting the off-diagonal blocks of $\Sigma$ and approximating each within-frequency pair by its asymptotic exponential law. The constructions \eqref{eq:block_lifted_like}--\eqref{eq:block_rho_S} therefore provide a principled interpolation between such fully factorized likelihoods and the exact correlated finite-$T$ theory.

\begin{figure}[H]
\centering
\includegraphics[width=8cm,height=7cm]{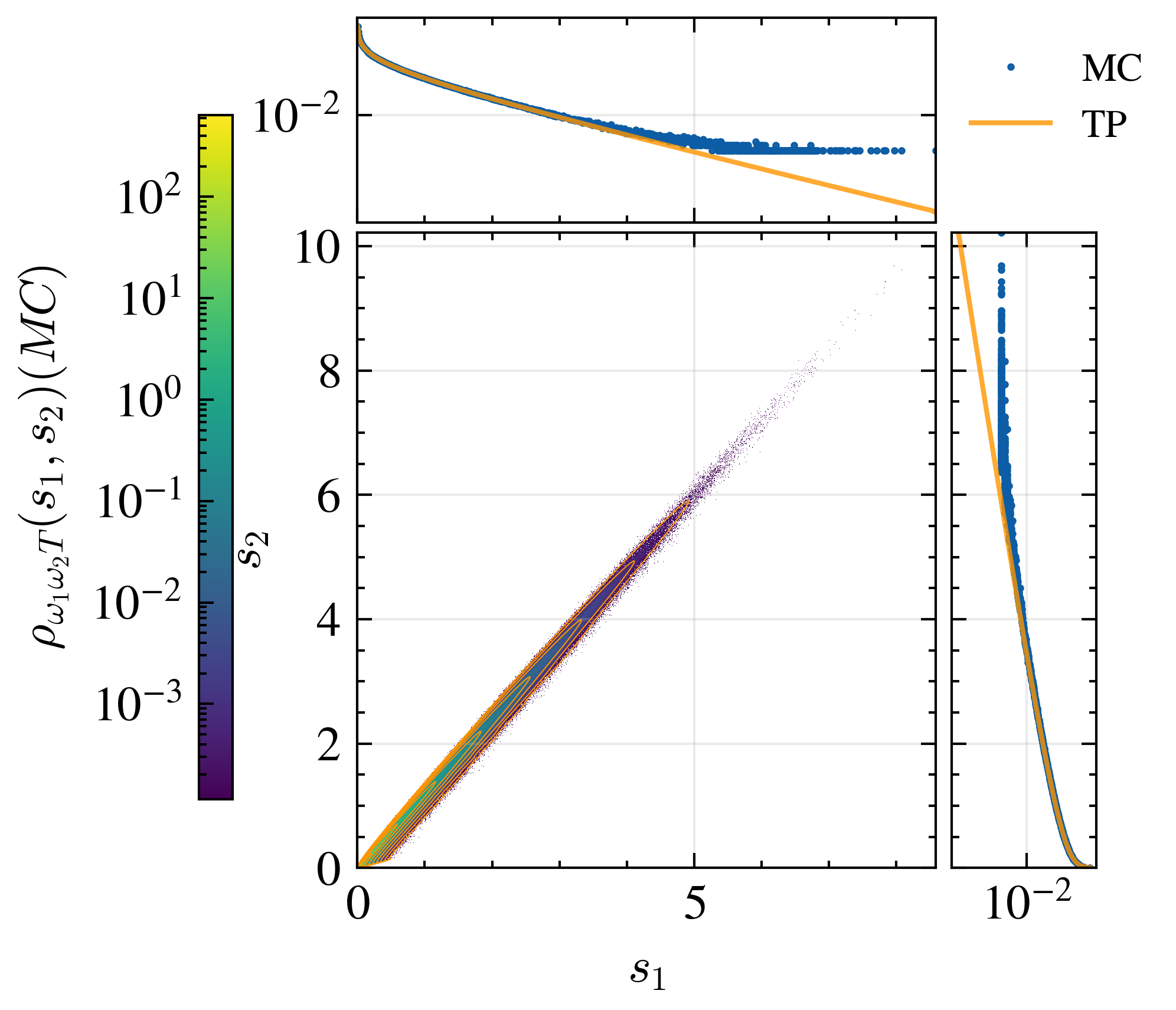}
\includegraphics[width=8cm,height=7cm]{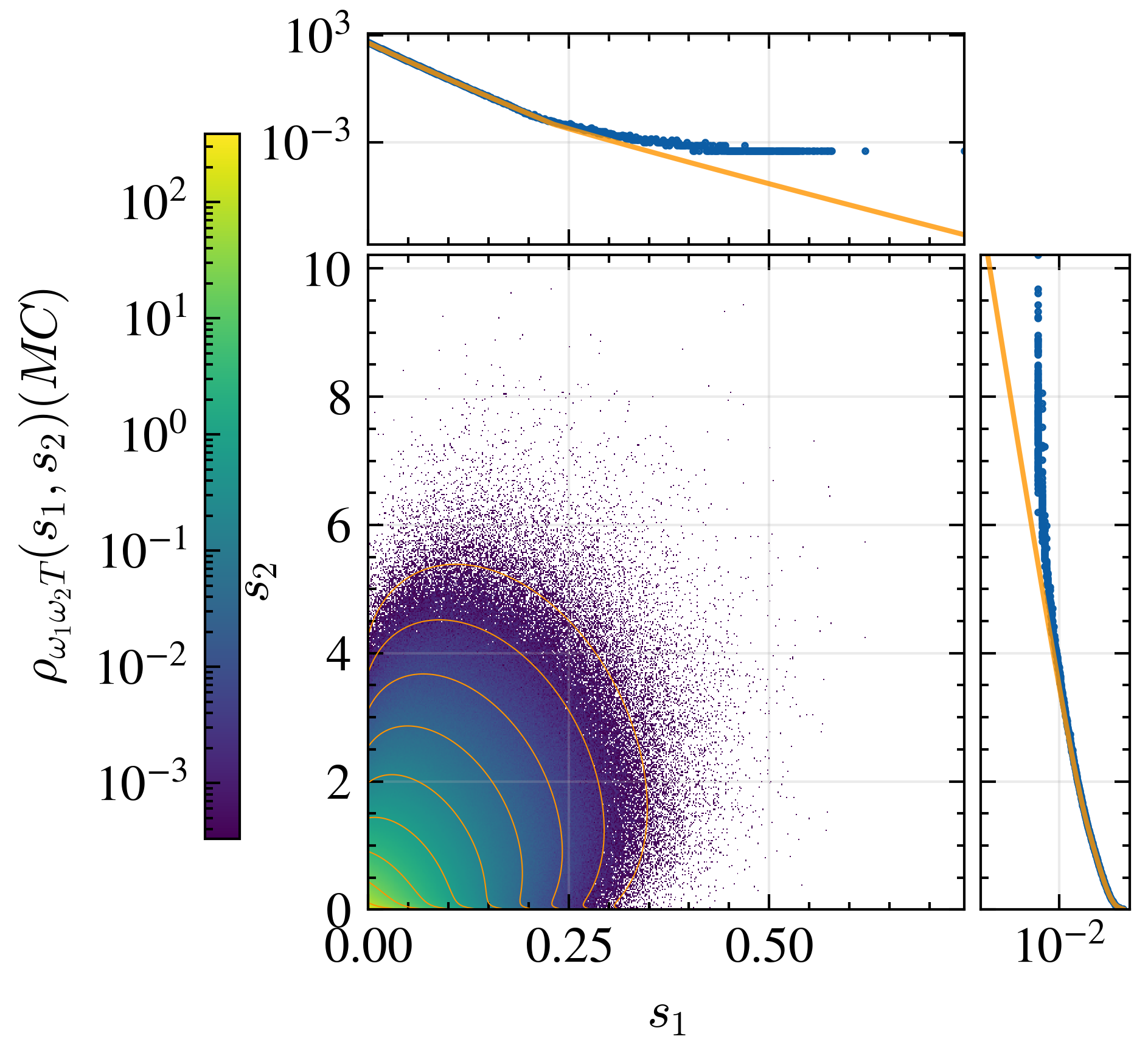}
\includegraphics[width=8cm,height=7cm]{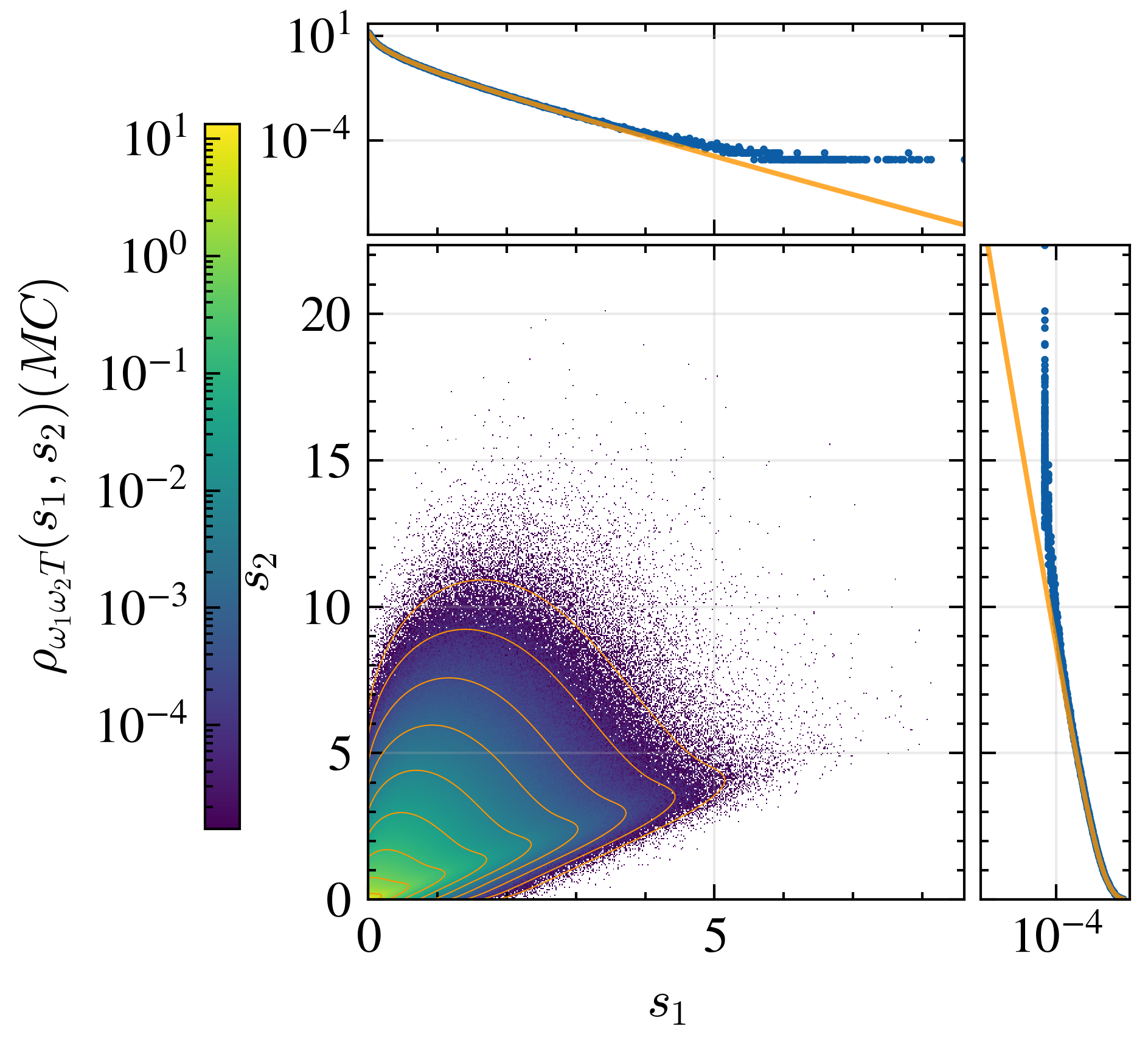}
\includegraphics[width=8cm,height=7cm]{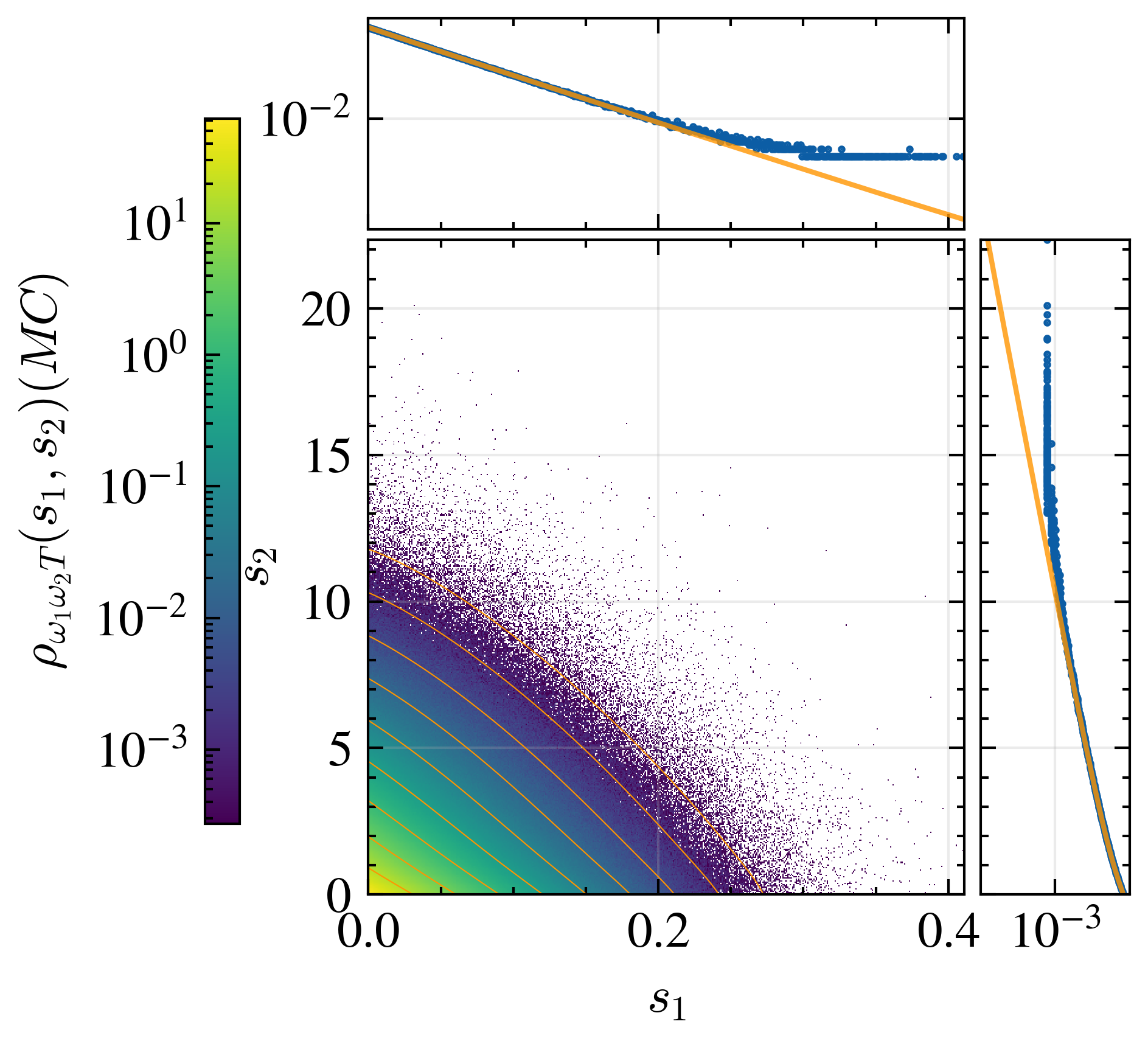}
\caption{Joint probability density function $\rho_{\omega_{1}\omega_{2},T}(s_{1},s_{2})$ in the $(s_1,s_2)$-plane. The color map shows Monte Carlo (MC) results ($10^8$ samples), while the orange curves show the theoretical prediction (TP). Upper panels: observation time $T=1$; lower panels: $T=3$. Left panels: $\omega_1 = 2$; right panels: $\omega_1 = 10$. Other parameters are fixed at $\omega_2=1$, $\tau=1$, and $D=1$. The marginal projections for $s_1$ and $s_2$ are shown along the corresponding axes.}
\label{fig:jPDF}
\end{figure}

For completeness, one may also write the exact joint density directly in the original spectral variables $\{S(\omega_i,T)\}_{i=1}^L$.  After some algebra (see Appendix~\ref{appB}), one obtains
\beeq{
\rho_{\bm{\omega},T}(\bm{s}) &=\int_{-\infty}^{\infty} \left[\prod_{i=1}^L \frac{dv_idw_i}{2\pi}\right]\exp\Bigg[-\frac{2D}{T}\sum_{i,j=1}^L \Big( A^{(v_iv_j)}_{\omega_i,\omega_j,T} v_i v_j+A^{(w_iw_j)}_{\omega_i,\omega_j,T} w_i w_j+ A^{(v_iw_j)}_{\omega_i,\omega_j,T} v_i w_j+A^{(w_iv_j)}_{\omega_i,\omega_j,T} w_i v_j\Big) \Bigg]\\
&\times\prod_{i=1}^L J_0\left(\sqrt{2s_i\left(v_i^2+w_i^2\right)}\right)\,,
\label{eq:fulljoint}
}
where $J_0(x)$ denotes the Bessel function of the first kind. Although exact, the integrand in Eq.~\eqref{eq:fulljoint} is oscillatory and does not look like making the positivity of the resulting PDF manifest after integration. This in turn, makes the numerical estimation of the joint PDF \eqref{eq:fulljoint} a rather difficult task. Fortunately, the expression for $\rho_{\bm{\omega},T}(\bm{s}) $ can be rewritten in the explicitly positive form (details can be found in Appendix~\ref{appC})
\beeq{
\rho_{\bm{\omega},T}(\bm{s}) = \left(\frac{T}{8 \pi D}\right)^L \frac{1}{\sqrt{\det\mathcal{A}}} \int_{0}^{2\pi} \left[\prod_{i=1}^Ld\phi_i\right]\exp\left[ -\sum_{k,l=1}^L\sum_{\sigma,\tau=0,1} A'^{(\xi_k \xi_l)}_{\omega_k,\omega_l,T}\cos(\phi_k-\sigma\pi/2)\cos(\phi_l-\tau\pi/2)\right]\,,
}
where $\xi_i=v_i$ if $\sigma=0$ and $\xi_i=w_i$ if $\sigma=1$, and
\beeq{
A'^{(\xi_k \xi_l)}_{\omega_k,\omega_l,T}\equiv \frac{T}{4D}\sqrt{s_ks_l}\,[\mathcal{A}^{-1}]^{(\xi_k \xi_l)}_{\omega_k,\omega_l,T}\,.
}
Figure~\ref{fig:jPDF} illustrates the resulting exact joint density for two representative frequency pairs. In this direct representation the finite-$T$ multispectral couplings are visible at the level of the variables $\{S(\omega_i,T)\}$ themselves, while the Gaussian formulation \eqref{eq:liftedGaussian_multi} makes the same structure more transparent for inference.

%%%%%%%%%%%%%%%%%%%%%%%%%%%%%%%%%%%%%%%%%%%%%%%%%%%%%%%%%%%%%%%%%%%%%%%%%%%%%%
\section{Single-trajectory inference via hierarchical frequency-domain likelihoods}
\label{sec:inference}
%%%%%%%%%%%%%%%%%%%%%%%%%%%%%%%%%%%%%%%%%%%%%%%%%%%%%%%%%%%%%%%%%%%%%%%%%%%%%%
The finite-$T$ multispectral theory developed in Section~\ref{multiplefrequency} provides a direct route from \emph{single-trajectory} data to likelihoods in the frequency domain that retain, to varying degrees, the dependence structure induced by finite observation windows. For the sampled OU/AR(1) benchmark considered here, an exact time-domain Gaussian likelihood is available in principle. Accordingly, the purpose of this section is not to claim a fundamentally new optimal estimator for the OU model, but rather to use this exactly solvable setting as a controlled benchmark for the misspecification introduced by frequency-domain factorizations. The central question is then how single-trajectory inference changes as one moves from the exact likelihood of the full sampled process to retained-band spectral likelihoods of decreasing fidelity.

We consider a uniformly sampled trajectory $\{x_n\}_{n=0}^{N-1}$ with sampling interval $\Delta t$ and total duration $T=N\Delta t$. We work on the natural discrete Fourier grid
\beeq{
\omega_k=\frac{2\pi k}{T}\,,\qquad k=0,1,\ldots,\Big\lfloor\frac{N}{2}\Big\rfloor\,,
}
and define the discrete Fourier transform (DFT)
\beeq{
\widetilde x(\omega_k)\equiv \Delta t\sum_{n=0}^{N-1} x_n e^{-i\omega_k n\Delta t}\,.
\label{eq:disc_DFT}
}
For inference we retain a set of $L$ positive Fourier bins $\mathcal K\subset\{1,\ldots,\lfloor N/2\rfloor\}$ so that the corresponding periodogram ordinates are
\beeq{
I(\omega_k)\equiv \frac{1}{T}\left|\widetilde x(\omega_k)\right|^2\,,\qquad k\in\mathcal K\,.
\label{eq:disc_periodogram}
}
As a concrete parametric Gaussian model we use the exact discretization of the continuous-time OU process,
\beeq{
x_{n+1}=a x_n+\sigma\varepsilon_n\,, \qquad \varepsilon_n\sim\mathcal{N}(0,1)\,,\qquad a=\exp(-\Delta t/\tau)\,,
\label{eq:disc_OU_AR1}
}
with noise amplitude chosen so that the parameters $(D,\tau)$ retain their continuous-time meaning,
\beeq{
\sigma^2 = D\tau(1-a^2)\,.
\label{eq:disc_OU_sigma}
}
Throughout, we continue considering the deterministic initial condition (in particular $x_0=0$) and treat $(D,\tau)$ as unknown.

For the sampled OU/AR(1) model \eqref{eq:disc_OU_AR1}-- \eqref{eq:disc_OU_sigma}, the exact conditional Gaussian log-likelihood of the full trajectory is
\beeq{
\ell_{\rm time}(D,\tau) = -\frac{N-1}{2}\log\big(D\tau(1-a^2)\big) -\frac{1}{2D\tau(1-a^2)}\sum_{n=0}^{N-2}(x_{n+1}-a x_n)^2+\text{const}\,.
}
Its maximizer $(\widehat D_{\rm time},\widehat\tau_{\rm time})=\arg\max_{D,\tau}\ell_{\rm time}(D,\tau)$ is the exact OU benchmark under the assumed data-generating model. If instead one works with a retained spectral summary, then the Gaussian representation derived in Section~\ref{multiplefrequency}  gives a second exact object. Writing
\beeq{
\widetilde x(\omega_k)=Z_{c,k}-iZ_{s,k}\,,\qquad Z_{c,k}=\Re\widetilde x(\omega_k)\,,\qquad Z_{s,k}=-\Im\widetilde x(\omega_k)\,,
\label{eq:Zcs_from_DFT}
}
so that $I(\omega_k)=(Z_{c,k}^2+Z_{s,k}^2)/T$ for $k\in\mathcal K$, we collect the real Fourier coefficients into the vector
\beeq{
\bm{z}\equiv\Big(\{Z_{c,k}\}_{k\in\mathcal K},\{Z_{s,k}\}_{k\in\mathcal K}\Big)^{\top}\in\mathbb R^{2L}\,.
\label{eq:z_lifted_vector_disc}
}
For Gaussian dynamics, $\bm z$ is exactly multivariate normal at finite $T$,
\beeq{
\bm z\sim \mathcal N\big(0,\Sigma(D,\tau)\big)\,,\qquad \Sigma(D,\tau)=D\Sigma_1(\tau)\,,
\label{eq:z_gaussian_disc}
}
where $\Sigma_1(\tau)$ is the $2L\times 2L$ lifted covariance at unit $D$. Hence the exact finite-$T$ likelihood on the retained spectral variables is
\beeq{
\ell_{\rm spec}^{\rm full}(D,\tau)=-\frac{1}{2}\log\det\Sigma(D,\tau)-\frac{1}{2}\bm{z}^\top\Sigma(D,\tau)^{-1}\bm{z}+\text{const}\,.
}
Notice that this likelihood is indeed exact for the chosen spectral representation and coincides with the fully correlated endpoint of the Gaussian block hierarchy introduced in Eq. \eqref{eq:block_lifted_like}. It is, however, already different from $\ell_{\rm time}(D,\tau)$ because it uses only the retained spectral variables for a subsect of frequencies rather than the full trajectory. Our inferential benchmark therefore contains two logically distinct losses: first passing from full data to a retained spectral summary and second simplifying the dependence structure inside the spectral summary.

More concretely for this second loss: we consider covariance truncations of the exact retained-band spectral likelihood. We partition the index set $\mathcal K$ into disjoint blocks
\beeq{
\mathcal K=\mathcal{B}_1\cup\cdots\cup \mathcal{B}_M\,, \qquad \mathcal{B}_\alpha\cap \mathcal{B}_\beta=\varnothing (\alpha\neq\beta)\,,
\label{eq:block_partition}
}
typically as contiguous blocks of fixed size $m$ in the ordered list of Fourier indices (with perhaps the last block being smaller). For each block $\mathcal{B}_{\alpha}$ we define the corresponding subvector $\bm{z}_{\mathcal{B}_{\alpha}}$ and the corresponding principal submatrix $\Sigma_{\mathcal{B}_{\alpha}}(D,\tau)$. The block-Gaussian likelihood is then
\beeq{
\mathcal L_{\rm blk}^{(\{\mathcal B_\alpha\})}(D,\tau\mid \{\bm z_{\mathcal B_\alpha}\}) \equiv \prod_{\alpha=1}^{M} \frac{1}{(2\pi)^{|\mathcal B_\alpha|}\sqrt{\det \Sigma_{\mathcal B_\alpha}(D,\tau)}} \exp\left[-\frac{1}{2}\bm z_{\mathcal B_\alpha}^{T}[\Sigma_{\mathcal B_\alpha}(D,\tau)]^{-1}\bm z_{\mathcal B_\alpha}\right]\,.
\label{eq:block_lifted_like_V}
}
Up to an additive constant, the corresponding log-likelihood is
\beeq{
\ell_{\rm blk}^{(\{\mathcal B_\alpha\})}(D,\tau)=-\frac{1}{2}\sum_{\alpha=1}^{M}\left[\log\det \Sigma_{\mathcal B_\alpha}(D,\tau)+\bm z_{\mathcal B_\alpha}^{T}[\Sigma_{\mathcal B_\alpha}(D,\tau)]^{-1}\bm z_{\mathcal B_\alpha}\right]\,.
\label{eq:blk_loglik}
}
Using $\Sigma_{\mathcal{B}_\alpha}(D,1)=D\Sigma_{1,\mathcal{B}_{\alpha}}(\tau)$ and $n\equiv 2L$, the maximization over $D$ can again be done explicitly:
\beeq{
\widehat{D}_{\rm blk}(\tau)=\frac{Q(\tau)}{n}\,, \qquad Q(\tau)\equiv\sum_{\alpha=1}^{M}\bm z_{\mathcal B_\alpha}^{\top}\Sigma_{1,\mathcal B_\alpha}(\tau)^{-1}\bm z_{\mathcal B_\alpha}\,.
\label{eq:Dhat_block_profile}
}
Substituting \eqref{eq:Dhat_block_profile} into \eqref{eq:blk_loglik} gives the cost function for $\tau$,
\beeq{
\ell_{\rm blk,prof}(\tau)=-\frac{1}{2}\sum_{\alpha=1}^{M}\log\det\Sigma_{1,\mathcal B_\alpha}(\tau)-\frac{n}{2}\log\!\left(\frac{Q(\tau)}{n}\right)+\mathrm{const.}\,,
\label{eq:tau_profile_block}
}
which we maximize numerically, and then set $\widehat{D}_{\rm blk}=\widehat{D}_{\rm blk}(\widehat{\tau}_{\rm blk})$. This yields a monotone hierarchy at the covariance level where  $m=1$ corresponds to a per-frequency Gaussian treatment of the Fourier coefficient pairs, while for $m=L$ recovers the fully correlated Gaussian likelihood on the selected frequency band.

Alongside this Gaussian hierarchy we also consider two scalar-ordinate models built directly on the periodograms $\{I(\omega_k)\}_{k\in\mathcal K}$. A first finite-$T$ refinement keeps the factorization across frequency bins but replaces each asymptotic exponential factor by the exact single-bin finite-$T$ law. Denoting by $\rho^{\rm FT}_{\omega_k,T}(s\mid D,\tau)$ the exact one-frequency density of $I(\omega_k)$ for the sampled OU model given by Eq.~\eqref{eq:pdffinal}, the corresponding factorized log-likelihood is
\beeq{
\ell_{\rm FT}(D,\tau)=\sum_{k\in\mathcal K}\log \rho^{\rm FT}_{\omega_k,T}\big(I(\omega_k)\mid D,\tau\big)\,.
}
This likelihood corrects the within-frequency finite-$T$ law while still neglecting all cross-frequency dependence. In the present Gaussian setting, each factor is determined by the corresponding $2\times 2$ covariance of the Fourier pair $(Z_{c,k},Z_{s,k})$ and is therefore a finite-$T$ Bessel-type single-bin density rather than a asymptotic exponential approximation.

The coarsest frequency-domain approximation is the classical Whittle model \cite{Whittle1953}, which treats the retained periodogram ordinates as approximately independent exponential variables with means given by the model spectral density. For the discrete OU/AR(1) model \eqref{eq:disc_OU_AR1}--\eqref{eq:disc_OU_sigma}, with the normalization \eqref{eq:disc_DFT}--\eqref{eq:disc_periodogram}, the Whittle mean can be written in separable form
\beeq{
\mathbb{E} I(\omega_k)\approx S_{D,\tau}(\omega_k)=D g(\omega_k;\tau)\,, \qquad g(\omega;\tau)=\frac{\Delta t \tau (1-a^2)}{1-2a\cos(\omega\Delta t)+a^2}\,, \qquad a=e^{-\Delta t/\tau}\,.
\label{eq:whittle_mean_discrete_OU}
}
The corresponding Whittle log-likelihood reads
\beeq{
\ell_{\rm W}(D,\tau)=-\sum_{k\in\mathcal{K}}\left[\log \big(D g(\omega_k;\tau)\big) +\frac{I(\omega_k)}{D\,g(\omega_k;\tau)}\right]+\mathrm{const}\,.
\label{eq:whittle_loglik}
}
Because $S_{D,\tau}(\omega)=D g(\omega;\tau)$ is linear in $D$, maximization over $D$ can be profiled analytically at fixed $\tau$:
\beeq{
\widehat D_{\rm W}(\tau) = \frac{1}{L} \sum_{k\in\mathcal{K}} \frac{I(\omega_k)}{g(\omega_k;\tau)}\,.
\label{eq:Dhat_whittle_profile}
}
Substituting \eqref{eq:Dhat_whittle_profile} into \eqref{eq:whittle_loglik} yields a one-dimensional profile objective for $\tau$, which we maximize numerically, and then set $\widehat D_{\rm W}=\widehat D_{\rm W}(\widehat\tau_{\rm W})$.

It is important to distinguish these spectral likelihoods carefully. The finite-$T$ factorized model and the Whittle model act  on the scalar ordinates $\{I(\omega_k)\}$, differing solely in whether the  one-bin law is exact at finite $T$ or replaced by its asymptotic exponential approximation. By contrast, the $m=1$ Gaussian model acts on the two-dimensional Fourier pairs $\{(Z_{c,k},Z_{s,k})\}$ and treats them as independent across $k$ but not necessarily isotropic within each frequency. In the asymptotic regime where each Fourier pair is approximately circular Gaussian, the $m=1$ Gaussian model and the Whittle model nearly coincide; at finite $T$, however, they do not need to, because within-frequency anisotropy and cross-frequency leakage are both present.

We assess this hierarchy by generating $n_{\rm traj}$ independent discrete-time OU trajectories from Eqs. \eqref{eq:disc_OU_AR1}--\eqref{eq:disc_OU_sigma} at fixed ground-truth parameters $(\tau_{\star},D_{\star})$ and fixed observation protocol $(T,\Delta t)$, extracting the periodogram and the Fourier vector over the same retained band $\mathcal{K}$ for each trajectory, and computing the  maximum likelihood estimators under the spectral likelihoods described above. Because an exact time-domain Gaussian likelihood exists for this benchmark, the results below should be read as a controlled comparison between frequency-domain pseudo-likelihoods of increasing fidelity, not as a claim that spectral methods outperform exact OU maximum likelihoods estimators.

\begin{table}[t]
\centering
\begin{tabular}{l S S S S S S}
\toprule
{Method}
& {$\mathrm{median}(\hat{\tau})$}
& {$\mathrm{IQR}(\hat{\tau})$}
& {$\mathrm{RMSE}(\hat{\tau})$}
& {$\mathrm{median}(\hat{D})$}
& {$\mathrm{IQR}(\hat{D})$}
& {$\mathrm{RMSE}(\hat{D})$} \\
\midrule
Whittle                         & 9.034 & 3.427 & 3.296 & 2.074 & 0.224 & 0.245 \\
Finite-$T$ factorized           & 9.440 & 3.776 & 3.623 & 2.026 & 0.245 & 0.236 \\
Gaussian ($m=1$)            & 9.355 & 3.950 & 3.658 & 2.028 & 0.237 & 0.233 \\
Gaussian ($m=10$)           & 9.800 & 4.637 & 8.837 & 2.008 & 0.255 & 0.208 \\
Gaussian ($m=20$)           & 9.860 & 5.046 & 5.980 & 2.011 & 0.250 & 0.206 \\
Gaussian ($m=100$)          & 9.620 & 4.001 & 3.589 & 2.011 & 0.254 & 0.210 \\
\bottomrule
\end{tabular}
\caption{Robust and accuracy summaries of per-trajectory MLE estimates for the OU parameters $(\tau,D)$ using the Whittle likelihood, the finite-$T$ factorized likelihood, and Gaussian likelihood approximations at representative block sizes. Settings: $T=200$, $\Delta t=10^{-3}$, $L=100$ retained positive Fourier bins, $N_{\mathrm{traj}}=100$, with ground-truth parameters $(\tau_\star,D_\star)=(10,2)$. The interquartile range (IQR) is $q_{0.75}-q_{0.25}$.}
\label{tab:main_mle_T200_dt1e-3}
\end{table}

Table~\ref{tab:main_mle_T200_dt1e-3} should be read as a broad-band misspecification benchmark. The cleanest pattern is the strong separation in the inferential process between the two parameters. The diffusion coefficient is recovered fairly robustly across all approximations: $\mathrm{median}(\hat{D})$ remains close to $D_{\star}=2$, with relative narrow IQR and small RMSE. By contrast, inference  of the relaxation time from a single finite record is much more fragile. Even when the median remains near $\tau_{\star}=10$, the spread in $\hat{\tau}$ is several units wide, so the inferential cost of the likelihood approximation is felt more strongly in $\tau$ than in $D$.

The second important observation is that restoring cross-frequency dependence is not monotone in performance. The Gaussian family changes the behavior of the $\tau$ estimator in a non trivial way: the median and IQR evolve only moderately with $m$, but the $\tau$ RMSE can deteriorate sharply at intermediate block sizes. In particular, the large value of $\mathrm{RMSE}(\hat{\tau})$ at $m=10$ is not accompanied by a commensurate shift in the median, which indicates that the squared-error increase is driven by relatively rare but severe mis-estimation events. By $m=100$, the $\tau$-RMSE returns to scale of the factorized baselines. Comparison between Whittle and the finite-$T$ factorized model isolates the effect of correcting the within-frequency law while still ignoring all inter-frequency coupling, whereas comparison between the finite-$T$ factorized model and the Gaussian family isolates the additional effect of restoring cross-frequency covariance. In the present benchmark these two effects are qualitatively different: the former produces relatively mild shifts in robust summaries, whereas the latter can substantially modify the tail behavior of $\hat{\tau}$.

These broad-band summaries already show that the finite-$T$ misspecification affects single-trajectory inference in a structured and parameter-dependent way, but they do not yet calibrate nominal confidence sets or predictive scores against the exact benchmark. To sharpen that point we now keep the same sampled OU/AR(1) model and the same spectral hierarchy, but restrict our attention to a short-record, low-frequency stress test anchored in the exact time-domain Gaussian likelihood. Concretely, we work on the low contiguous band $\mathcal{K}=\{1,\ldots,L_{\rm low}\}$, where $L_{\rm low}$ denotes the number of retained positive Fourier bins, so that e.g. $L_{\rm low}=20$ means that only the first 20 positive Fourier frequencies are used in the fit.

We summarize this calibration benchmark through four diagnostics for $\tau$. The first is the empirical coverage of the nominal $95\%$ profile-likelihood interval, denoted here as $\mathrm{cover}_{0.95}(\tau)$. The second is the large-error rate
\beeq{
p_{0.5}=\text{Prob}\left(\left|\hat{\tau}-\tau_{\star}\right|> 0.5\tau_{\star}\right)\,,
}
which is the fraction of trajectories for which the inferred relaxation time differs from the true value by more than $50\%$. The third is the mean exact-time score gap per transition. Denoting as  $\widehat{\theta}_{\rm time}=(\widehat{D}_{\rm time},\widehat{\tau}_{\rm time})$ the exact benchmark estimate and as $\widehat{\theta}_{\rm meth}$ the estimated delivered by any spectral method, we define
\beeq{
\Delta \ell_{\rm step}=\frac{\ell_{\rm time}(\widehat{\theta}_{\rm time})-\ell_{\rm time}(\widehat{\theta}_{\rm meth})}{N-1}\,,
}
Since $\widehat{\theta}_{\rm time}$ maximizes the exact time-domain likelihood, $\Delta \ell_{\rm step}\ge 0$. Small values therefore mean that the spectral estimator is close to the exact benchmark in likelihood terms. The fourth diagnostic, used only for the block Gaussian family, is the covariance-truncation ratio
\beeq{
E(m)\equiv\frac{\|\Sigma-\Sigma^{(m)} \|_{F}}{\|\Sigma\|_{F}}\,.
}
This quantity measures the fraction of the covariance discarded when the full covariance $\Sigma$ is replaced by its block-diagonal approximation $\Sigma^{(m)}$.

These fourth diagnostic tools separate the two losses already identified above. Even at the fully correlated endpoint $m=L_{\rm low}$ one no longer uses the full trajectory likelihood, but only the retained low-frequency spectral summary. Moving from $m=L_{\rm low}$ to smaller $m$, or from the Gaussian family to the finite-$T$ factorized or Whittle models, deletes additional finite-$T$ spectral dependence. The calibration benchmark is designed so that comparisons among Whittle, finite-$T$ factorized, and block Gaussian methods isolate this second effect, while the comparison with the exact time-domain benchmark fixes the overall scale of inferential deterioration.

\begin{table}[t]
\centering
\small
\begin{tabular}{lcccc}
\toprule
Method & $\mathrm{cover}_{0.95}(\tau)$ & $p_{0.5}$ & $\Delta \ell_{\rm step}$ & $E(m)$ \\
\midrule
Time exact                & 0.996000 & 0.609667 & 0.000000 & -- \\
Whittle                   & 1.000000 & 0.878000 & 0.028313 & -- \\
Finite-$T$ factorized     & 1.000000 & 0.720667 & 0.033632 & -- \\
Block-lifted ($m=1$)      & 0.987000 & 0.720667 & 0.022226 & 0.324837 \\
Block-lifted ($m=2$)      & 0.941000 & 0.666000 & 0.016784 & 0.263355 \\
Block-lifted ($m=3$)      & 0.943667 & 0.672333 & 0.014972 & 0.223048 \\
Block-lifted ($m=4$)      & 0.947000 & 0.650667 & 0.014167 & 0.194087 \\
Block-lifted ($m=5$)      & 0.958333 & 0.649667 & 0.013333 & 0.171830 \\
Block-lifted ($m=8$)      & 0.968333 & 0.629667 & 0.012911 & 0.125890 \\
Block-lifted ($m=10$)     & 0.983333 & 0.625667 & 0.012290 & 0.104068 \\
Block-lifted ($m=15$)     & 0.985667 & 0.627000 & 0.012208 & 0.061509 \\
Block-lifted ($m=20$)     & 0.998000 & 0.623667 & 0.011829 & 0.000000 \\
\bottomrule
\end{tabular}
\caption{Calibration-focused short-record benchmark for $T/\tau_{\star}=2$ and $L_{\rm low}=20$, i.e. using the first 20 positive Fourier bins. The first column reports the empirical coverage $\mathrm{cover}_{0.95}(\tau)$ of the nominal $95\%$ profile-likelihood interval, $p_{0.5}\equiv \Pr(|\hat\tau-\tau_{\star}|>0.5\tau_{\star})$ is the large-error rate, $\Delta \ell_{\rm step}$ denotes the mean exact-time score gap per transition relative to the exact time-domain Gaussian benchmark, and $E(m)\equiv \|\Sigma-\Sigma^{(m)}\|_F/\|\Sigma\|_F$.}
\label{tab:short_record_calibration}
\end{table}

Table~\ref{tab:short_record_calibration} makes the calibration picture considerably sharper than Table~\ref{tab:main_mle_T200_dt1e-3}. The coverage column is informative but not the main discriminator: it remains uniformly high, ranging from $0.941000$ to $1.000000$, with the exact benchmark itself at $0.996000$. The sharper separation appears in $p_{0.5}$ and $\Delta \ell_{\rm step}$. Relative to the exact benchmark, the Whittle approximation raises $p_{0.5}$ from $0.609667$ to $0.878000$ and incurs $\Delta \ell_{\rm step}=0.028313$. The finite-$T$ factorized model lowers the tail probability to $0.720667$, but its exact-time score gap is actually larger, $\Delta \ell_{\rm step}=0.033632$, showing that correcting the one-bin law alone is not sufficient. By contrast, progressively restoring cross-frequency covariance through the block-lifted hierarchy yields a near-monotone reduction in both diagnostics: from $m=1$ to $m=20$, $p_{0.5}$ falls from $0.720667$ to $0.623667$, while $\Delta \ell_{\rm step}$ falls from $0.022226$ to $0.011829$. Over the same range, $E(m)$ decreases from $0.324837$ to $0$, which is precisely the theorem-driven pattern suggested by Section~\ref{multiplefrequency}. As an auxiliary control, the diffusion parameter remains comparatively stable in this regime: the corresponding $D$-biases for the block-lifted family stay between $-0.012872$ and $0.027509$, whereas Whittle and the finite-$T$ factorized model give $0.504849$ and $0.112881$, respectively.

To test whether this ordering is specific to one stress-test regime, we next vary the record length through the ratio $T/\tau^\ast$ while keeping the same low contiguous band structure. Since the retained set is always chosen as $\mathcal K=\{1,\ldots,L_{\rm low}\}$ in index space, changing $T$ also changes the physical frequencies $\omega_k=2\pi k/T$; the sweep should therefore be read as a sequence of finite-$T$ stress tests rather than as a pure asymptotic scaling study. Table~\ref{tab:short_record_sweep} summarizes the results for $L_{\rm low}=20$, which is the more informative low-frequency band because it allows us to compare moderate block sizes with the fully correlated endpoint $m=L_{\rm low}$. The stress-test table below is based on an independent Monte Carlo run and is intended as a robustness check, so its entries should be compared qualitatively with Table~\ref{tab:short_record_calibration} rather than numerically line by line.

\begin{table*}[t]
\centering
\small
\setlength{\tabcolsep}{5pt}
\begin{tabular}{@{}c c c c c c@{}}
\toprule
$T/\tau^\ast$ & Time exact & Whittle & \shortstack{Finite-$T$\\fact.} & \shortstack{Block\\$m=10$} & \shortstack{Block\\$m=20$} \\
\cmidrule(lr){1-6}
\multicolumn{6}{c}{$p_{0.5}$ for $L_{\rm low}=20$} \\
\cmidrule(lr){1-6}
$2$  & 0.627 & 0.876 & 0.741 & 0.639 & 0.648 \\
$3$  & 0.537 & 0.776 & 0.685 & 0.586 & 0.583 \\
$5$  & 0.372 & 0.595 & 0.585 & 0.520 & 0.483 \\
$10$ & 0.223 & 0.364 & 0.374 & 0.359 & 0.355 \\
$20$ & 0.104 & 0.216 & 0.244 & 0.247 & 0.243 \\
\addlinespace[0.6ex]
\cmidrule(lr){1-6}
\multicolumn{6}{c}{$\Delta \ell_{\rm step}$ for $L_{\rm low}=20$} \\
\cmidrule(lr){1-6}
$2$  & 0.000000 & 0.027497 & 0.030607 & 0.011510 & 0.011082 \\
$3$  & 0.000000 & 0.022771 & 0.026582 & 0.012639 & 0.012047 \\
$5$  & 0.000000 & 0.017842 & 0.022611 & 0.014774 & 0.014328 \\
$10$ & 0.000000 & 0.018228 & 0.021056 & 0.017618 & 0.017005 \\
$20$ & 0.000000 & 0.025103 & 0.026622 & 0.025430 & 0.025396 \\
\bottomrule
\end{tabular}
\caption{Sensitivity sweep over $T/\tau^\ast$ for the low contiguous band $L_{\rm low}=20$, i.e. the first 20 positive Fourier bins. The upper block reports the large-error rate $p_{0.5}=\Pr(|\hat{\tau}-\tau^\ast|>0.5\tau^\ast)$, while the lower block reports the mean exact-time score gap per transition, $\Delta \ell_{\rm step}$. For the time-domain benchmark, $\Delta \ell_{\rm step}=0$ by definition.}
\label{tab:short_record_sweep}
\end{table*}

Two features stand out in Table~\ref{tab:short_record_sweep}. First, the short-record regimes $T/\tau_{\star}=2,3,5$ are the most discriminating. At $T/\tau_{\star}=2$, for example, $p_{0.5}$ is $0.876$ for Whittle and $0.741$ for the finite-$T$ factorized model, but drops to $0.639$ for block $m=10$ and to $0.648$ for the fully correlated endpoint $m=20$, against $0.627$ for the exact benchmark; the corresponding score gaps are $0.027497$, $0.030607$, $0.011510$, and $0.011082$. At $T/\tau_{\star}=3$ the same ordering persists, with $p_{0.5}$ equal to $0.776$, $0.685$, $0.586$, and $0.583$, respectively, and score gaps equal to $0.022771$, $0.026582$, $0.012639$, and $0.012047$. Second, the separation narrows as $T/\tau_{\star}$ increases, so the added covariance structure matters most in the genuinely finite-time regime. The same qualitative ordering is also visible for the narrower low-frequency band $L_{\rm low}=10$, that is, when only the first 10 positive Fourier bins are retained: at $T/\tau^\ast=2$, $p_{0.5}$ drops from $0.884$ for Whittle to $0.674$ for block $m=10$, while $\Delta \ell_{\rm step}$ drops from $0.040876$ to $0.028015$; at $T/\tau_{\star}=3$, the corresponding values are $0.806$ to $0.636$ and $0.038247$ to $0.033790$.

Taken together, Tables~\ref{tab:main_mle_T200_dt1e-3}--\ref{tab:short_record_sweep} support a two-layer interpretation of the numerical benchmark. The broad-band moderate-record analysis of Table~\ref{tab:main_mle_T200_dt1e-3} shows that finite-$T$ misspecification changes the inferential behavior of spectral pseudo-likelihoods in a nontrivial and non-monotone way. The short-record low-frequency stress tests then show more sharply that, in the regime where the covariance structure of Section~\ref{multiplefrequency} is most relevant, restoring cross-frequency dependence brings the spectral likelihood substantially closer to the exact time-domain benchmark, whereas correcting only the one-bin finite-$T$ law is not enough. In this sense, the exact finite-$T$ multispectral theory provides not only a distributional characterization of spectral estimators, but also a calibration benchmark for frequency-domain inference.

From a practical viewpoint, the theory suggests two tuning decisions. The first is the choice of retained band $\mathcal K$. A conservative starting point is to work on the natural Fourier grid and select a contiguous band that excludes very low frequencies, where finite-$T$ transients or detrending can dominate, and also avoids bins too close to the Nyquist limit, where discretization and measurement noise are most intrusive. The second is the block size $m$. Since Section~\ref{multiplefrequency} predicts that window-induced correlations are strongest between nearby frequencies and decay beyond the leakage scale $\Delta\omega\sim 2\pi/T$, a reasonable strategy is to increase $m$ gradually until the inferred parameters or the profile log-likelihood cease to change appreciably. The smallest block size at which such saturation occurs provides a pragmatic compromise between computational cost and fidelity to the finite-$T$ covariance structure.

Overall, the discrete OU benchmark supports the following inferential reading of the finite-$T$ multispectral theory. The fully factorized Whittle model is only the coarsest endpoint of a broader family of finite-$T$ spectral pseudo-likelihoods. Correcting the single-bin law and restoring local cross-frequency covariance do not simply generate a monotone sequence of ``better'' estimators; rather, they expose how sensitive single-trajectory spectral inference can be to finite-window misspecification and identify deleted cross-frequency dependence as the main source of loss in the short-record regime where the benchmark is most discriminating.

%%%%%%%%%%%%%%%%%%%%%%%%%%%%%%%%%%%%%%%%%%%%%%%%%%%
\section{Conclusions}
\label{conclusions}
%%%%%%%%%%%%%%%%%%%%%%%%%%%%%%%%%%%%%%%%%%%%%%%%%%%
We have developed an exact finite-$T$ multispectral framework for spectral analysis from a single trajectory of an Ornstein--Uhlenbeck process, physically equivalent to an overdamped Brownian particle in a harmonic trap. At the single-frequency level, this yields the full finite-time law of the estimator $S(\omega,T)$ together with its asymptotic reductions. More importantly, at the multispectral level we obtained an exact characterization of the joint statistics of $\{S(\omega_i,T)\}_{i=1}^L$, thereby making explicit the inter-frequency couplings generated by finite windowing and absent from asymptotic frequency-factorized descriptions. The central structural result is an exact Gaussian representation, in which the cosine and sine Fourier projections form a multivariate normal vector $Z\sim\mathcal N(0,\Sigma)$ and the multispectral estimators are recovered as quadratic forms $S(\omega_i,T)=(Z_{c,i}^2+Z_{s,i}^2)/T$. This provides a covariance-explicit finite-$T$ joint-law / likelihood framework that goes beyond single-frequency fluctuation results and beyond pairwise correlation diagnostics. These predictions are supported by Monte Carlo simulations at the level of the single-frequency law, the multispectral covariance structure, and representative joint two-frequency densities.

The inferential role of this theory is best understood in calibration terms. For the sampled OU benchmark studied here, exact time-domain Gaussian likelihoods are available in principle, so the purpose of the frequency-domain hierarchy is not to claim a fundamentally superior OU estimator, but to quantify the effect of finite-$T$ spectral misspecification. In this setting, the Whittle likelihood is the coarsest diagonal-in-frequency endpoint, the finite-$T$ factorized model corrects the within-frequency law while still neglecting cross-frequency dependence, and the Gaussian block constructions progressively restore the covariance structure predicted by the exact finite-$T$ theory. The numerical results show that the effect of these approximations is parameter-dependent and non-monotone, especially for $\tau$, thereby illustrating that finite-window cross-frequency couplings are not merely formal corrections but can materially affect the behavior of single-trajectory spectral pseudo-likelihoods. In this sense, the main contribution is not only an explicit finite-$T$ multispectral law, but also a principled benchmark for diagnosing what is lost when one replaces the exact window-induced dependence structure by frequency-wise factorization.

The present results place classical spectral fitting in a more precise probabilistic setting and suggest several natural extensions. On the applied side, the framework is directly relevant to short-record trapped-particle measurements, where finite acquisition time, sampling effects, and spectral leakage are unavoidable. On the methodological side, the covariance-explicit lifted representation offers a natural basis for structured approximations in frequency space, including blockwise or more general sparsified models guided by the finite-$T$ dependence pattern itself. It also provides an analytically controlled benchmark for structured spectral-domain approximations, including data-driven or machine-learning approaches that attempt to learn dependence across frequencies \cite{yu2021whittle}. More broadly, the same perspective should extend to incorporate   experimentally more realistic acquisition effects, and multivariate trapped dynamics, where cross-frequency and cross-component dependence become central from the outset.

\acknowledgements
I.P.C.\ thanks colleagues at the Instituto de Ciencias F\'{\i}sicas (UNAM, Cuernavaca) for their hospitality and support during a difficult period. In particular, he thanks Juan Carlos, Antonio, Luis, and Thomas. This work was initiated with support from Christof Jung Kohl's CONAHCYT project (No.\ 425854). The final stage of this work was supported by SECIHTI under research grant CBF-2025-I-3911.

\bibliographystyle{apsrev4-1}
\bibliography{biblio.bib}

\appendix

%%%%%%%%%%%%%%%%%%%%%%%%%%%%%%%%%%%%%%%%%%%%%%%%%%%
\section{Single-frequency case}
\label{appA}
%%%%%%%%%%%%%%%%%%%%%%%%%%%%%%%%%%%%%%%%%%%%%%%%%%%
For the centered deterministic initialization used in the main text, $X_0=0$, the formal solution of the stochastic equation \eqref{eq:OU} is
\beeq{
X(t)=\sqrt{2D}\int_0^t ds e^{-(t-s)/\tau}W_x(s)\,.
}
Starting from the definition of the Laplace transform of the finite-time spectral estimator, we have
\beeq{
\Phi_{\omega,T}(\lambda)&=\mathbb{E}_{\{X(t)\}}\left[e^{- \lambda S(\omega,T )}\right]\\
&=\mathbb{E}_{\{X(t)\}}\left\{\exp\left[ -\frac{\lambda }{T} \left(\int_0^T dt  \cos(\omega t)X(t)\right)^2 -\frac{\lambda }{T}\left( \int_0^T dt \sin( \omega t)X(t)\right)^2\right]\right\}\\
&=\int\frac{dv dw}{2\pi} \exp\left[-\frac{v^2}{2}-\frac{w^2}{2}\right]\mathbb{E}_{\{X(t)\}}\exp\left[i \sqrt{\frac{2\lambda }{T}} \int_0^T dt  \left(v\cos(\omega t)+w\sin(\omega t)\right)X(t) \right]\\
&=\int\frac{dv dw}{2\pi} \exp\left[-\frac{v^2}{2}-\frac{w^2}{2}\right]\mathbb{E}_{\{W_x(t)\}}\exp\left[2i \sqrt{\frac{\lambda D}{T}} \int_0^T dt  g(t|v,w,\omega )\int_0^t ds e^{-(t-s)/\tau}W_x(s) \right]\,,
}
where we introduced
\beeq{
g(t|v,w,\omega)\equiv v\cos(\omega t)+w\sin(\omega t)\,.
}
Interchanging the order of integration gives
\beeq{
 \int_0^T dt  g(t|v,w,\omega )\int_0^t ds e^{-(t-s)/\tau}W_x(s)&=\int_0^ Tds W_x(s)\int_s^T dt g(t|v,w,\omega)e^{-(t-s)/\tau}\\
 &=\int_0^T ds W_x(s)\mathcal{F}(s|v,w,\omega,T)\,,
}
with
\beeq{
\mathcal{F}(s|v,w,\omega,T)=\int_s^T dt g(t|v,w,\omega)e^{-(t-s)/\tau}\,.
}
The average over the noise history is then Gaussian:
\beeq{
&\mathbb{E}_{\{W_x(t)\}}\exp\left[2i \sqrt{\frac{\lambda D}{T}} \int_0^T ds W_x(s)\mathcal{F}(s|v,w,\omega,T) \right]\\
&=\frac{1}{\mathcal{Z}}\int \mathcal{D}W_x \exp\left[-\frac{1}{2}\int_0^T ds W_x^2(s)+2i \sqrt{\frac{\lambda D}{T}} \int_0^T ds W_x(s)\mathcal{F}(s|v,w,\omega,T) \right]\\
&=\frac{1}{\mathcal{Z}}\int \mathcal{D}W_x \exp\left[-\frac{1}{2}\int_0^T ds \left(W_x(s)-2i  \sqrt{\frac{\lambda D}{T}} \mathcal{F}(s|v,w,\omega,T)\right)^2-\lambda\frac{2 D}{T}\int_0^T ds \mathcal{F}^2(s|v,w,\omega,T) \right]\\
&= \exp\left[-\lambda\frac{2 D}{T}\int_0^T ds \mathcal{F}^2(s|v,w,\omega,T)\right]\,.
}
Therefore,
\beeq{
G(dv,dw)&\equiv \frac{dv dw}{2\pi}\exp\left[-\frac{v^2}{2}-\frac{w^2}{2}\right]\,,\\
\Phi_{\omega,T}(\lambda)&=\int G(dv,dw) \exp\left[-\lambda\frac{2 D}{T}\int_0^T ds \mathcal{F}^2(s|v,w,\omega,T)\right]\,.
\label{eq:c}
}
Since $\mathcal{F}$ is linear in $v$ and $w$, we can write
\beeq{
\int_0^T ds \mathcal{F}^2(s|v,w,\omega,T)\equiv A^{(vv)}_{\omega,T} v^2+ A^{(ww)}_{\omega,T} w^2+2 A^{(vw)}_{\omega,T } v w\,,
}
where
\beeq{
A^{(vv)}_{\omega,T}&=\int_0^T ds\left(\int_s^T dt \cos(\omega t) e^{-(t-s)/\tau}\right)^2\,,\\
A^{(ww)}_{\omega,T}&=\int_0^T ds\left(\int_s^T dt \sin(\omega t) e^{-(t-s)/\tau}\right)^2\,,\\
A^{(vw)}_{\omega,T}&=\int_0^T ds\left(\int_s^T dt \cos(\omega t) e^{-(t-s)/\tau}\right)\left(\int_s^T dt \sin(\omega t) e^{-(t-s)/\tau}\right)\,.
}
These integrals can be explicitly carried out, yielding
\beeq{
A^{(vv)}_{\omega,T}&=\frac{\tau ^2 e^{-\frac{2 T}{\tau }} }{4 \omega \left(\omega^2 \tau ^2+1\right)^2}\Bigg(e^{\frac{2 T}{\tau }} \left(\omega^3 \tau ^3+2 \omega^3 \tau ^2 T+\omega \tau  \left(\omega \tau  \sin (2 \omega T)-\left(\omega^2 \tau ^2+1\right) \cos (2 \omega T)\right)-5 \omega \tau +2 \omega T+\sin (2 \omega T)\right)\\
&-2 \omega \tau  (\cos (\omega T)-\omega \tau  \sin (\omega T))^2-8 \omega \tau  e^{T/\tau } (\omega \tau  \sin (\omega T)-\cos (\omega T))\Bigg)\,,\\
A^{(ww)}_{\omega,T}&=\frac{\tau ^2 e^{-\frac{2 T}{\tau }} }{4 \omega \left(\omega^2 \tau ^2+1\right)^2}\Bigg(e^{\frac{2 T}{\tau }} \left(\omega \left(\omega^2 \tau ^3+2 \omega^2 \tau ^2 T-\tau +2 T\right)-\left(\omega^2 \tau ^2+1\right) \sin (2 \omega T)+\omega \tau  \left(\omega^2 \tau ^2+1\right) \cos (2 \omega T)\right)\\
&-2 \omega \tau  (\omega \tau  \cos (\omega T)+\sin (\omega T))^2\Bigg)\,,\\
A^{(vw)}_{\omega,T}&=\frac{\tau ^2 e^{-\frac{2 T}{\tau }}}{4 \omega \left(\omega^2 \tau ^2+1\right)^2} \Bigg(\omega \tau  \left(\left(\omega^2 \tau ^2-1\right) \sin (2 \omega T)-2 \omega \tau  \cos (2 \omega T)\right)\\
&-e^{\frac{2 T}{\tau }} \left(\omega^2 \tau ^2+\left(\omega^2 \tau ^2+1\right) (\omega \tau  \sin (2 \omega T)+\cos (2 \omega T))-1\right)+4 \omega \tau  e^{T/\tau } (\omega \tau  \cos (\omega T)+\sin (\omega T))\Bigg)\,.
\label{eq:expressionsA}
}

It is convenient to assemble these coefficients into the matrix
\beeq{
\bm{A}_{\omega,T}=\begin{pmatrix}
A^{(vv)}_{\omega,T}&A^{(vw)}_{\omega,T}\\
A^{(vw)}_{\omega,T}&A^{(ww)}_{\omega,T}
\end{pmatrix}\,.
}
Then the Gaussian integrals over $v$ and $w$ in \eqref{eq:c} can be carried out explicitly:
\beeq{
\Phi_{\omega,T}(\lambda)&=\int G(dv,dw) \exp\left[-\lambda\frac{2 D}{T}\left(A^{(vv)}_{\omega,T} v^2+ A^{(ww)}_{\omega,T} w^2+2 A^{(vw)}_{\omega,T } v w\right)\right]\\
&=\int \frac{dv dw}{2\pi}\exp\left[-\frac{1}{2}\begin{pmatrix}v&w
\end{pmatrix}\left(\mathbb{I}+\lambda\frac{4D}{T}\bm{A}_{\omega,T}\right)\begin{pmatrix}v\\w\end{pmatrix}\right]\\
&=\left[\det\left(\mathbb{I}+\lambda\frac{4D}{T}\bm{A}_{\omega,T}\right)\right]^{-1/2}\\
&=\left[1+\lambda\frac{4D}{T} \left(A^{(vv)}_{\omega,T}+A^{(ww)}_{\omega,T}\right)+\lambda^2\left(\frac{4D}{T}\right)^2\left(A^{(vv)}_{\omega,T}A^{(ww)}_{\omega,T}-[A^{(vw)}_{\omega,T}]^2\right)\right]^{-1/2}\,.
\label{eq:MGF}
}
From here we can obtain the cumulants of the random variable $S(\omega,T)$. Since \eqref{eq:MGF} is a Laplace transform, the cumulant generating function is expanded as
\beeq{
\Psi_{\omega,T}(\lambda)&\equiv \log\Phi_{\omega,T}(\lambda)=-\frac{1}{2}\log\left[1+\lambda\mathcal{A}^{(1)}_{\omega,T}+\lambda^2 \mathcal{A}^{(2)}_{\omega,T}\right]
=\sum_{r=1}^\infty \frac{(-1)^r}{r!}\lambda^r \kappa_r\,,
}
with
\beeq{
\mathcal{A}^{(1)}_{\omega,T}&\equiv \frac{4D}{T} \left(A^{(vv)}_{\omega,T}+A^{(ww)}_{\omega,T}\right)\,,\\
\mathcal{A}^{(2)}_{\omega,T}&\equiv \left(\frac{4D}{T}\right)^2\left(A^{(vv)}_{\omega,T}A^{(ww)}_{\omega,T}-[A^{(vw)}_{\omega,T}]^2\right)\,.
}
Expanding the logarithm gives
\beeq{
\Psi_{\omega,T}(\lambda)&=\frac{1}{2}\sum_{n=1}^\infty \frac{(-1)^n}{n}\left(\lambda\mathcal{A}^{(1)}_{\omega,T}+\lambda^2 \mathcal{A}^{(2)}_{\omega,T}\right)^n\\
&=\frac{1}{2}\sum_{n=1}^\infty \frac{(-1)^n}{n}\sum_{\ell=0}^n\binom{n}{\ell} \lambda^{\ell+2(n-\ell)}\left(\mathcal{A}^{(1)}_{\omega,T}\right)^\ell\left( \mathcal{A}^{(2)}_{\omega,T}\right)^{n-\ell}\,,
}
and therefore
\beeq{
\kappa_r&=(-1)^r\frac{r!}{2}\sum_{n=1}^\infty \frac{(-1)^n}{n}\sum_{\ell=0}^n\binom{n}{\ell} \left(\mathcal{A}^{(1)}_{\omega,T}\right)^\ell\left( \mathcal{A}^{(2)}_{\omega,T}\right)^{n-\ell}\delta_{r,\ell+2(n-\ell)}\,.
}
In particular,
\beeq{
\kappa_1&=\frac{2D}{T} \left(A^{(vv)}_{\omega,T}+A^{(ww)}_{\omega,T}\right)\,,\\
\kappa_2&=\frac{1}{2}\left(\frac{4D}{T}\right)^2\left\{ [A^{(vv)}_{\omega,T}]^2+[A^{(ww)}_{\omega,T}]^2+2[A^{(vw)}_{\omega,T}]^2\right\}\,,
}
so that
\beeq{
\Phi_{\omega,T}(\lambda)=\left[1+2\lambda\kappa_1+\lambda^2\left(2\kappa_1^2-\kappa_2\right)\right]^{-1/2}\,.
}
This is the form used in Section~\ref{onefrequency}, where $\kappa_1=\mu_{\omega,T}$ and $\kappa_2=\sigma_{\omega,T}^2$.

It is also useful to record the corresponding formulas for a general Gaussian initial condition. If $X(0)=X_0$, then
\beeq{
X(t)=e^{-t/\tau}X_0+\sqrt{2D}\int_0^t ds e^{-(t-s)/\tau}W_x(s)\,.
}
Let $m_0=\mathbb{E}[X_0]$ and $V_0=\text{Var}(X_0)$. Assuming that $X_0$ is independent of the driving noise, one finds
\beeq{
\mathbb{E}[X(t)]=m_0e^{-t/\tau}\,,\quad\quad \mathbb{E}[X(t)X(t')]=D\tau e^{-|t-t'|/\tau}+(V_0+m_0^2-D\tau)e^{-(t+t')/\tau}\,,
}
so that the covariance becomes
\beeq{
C(t,t')\equiv\mathbb{E}[X(t)X(t')]-\mathbb{E}[X(t)]\mathbb{E}[X(t')]=D\tau e^{-|t-t'|/\tau}+(V_0-D\tau)e^{-(t+t')/\tau}\,.
}
In particular, the stationary Gaussian preparation is recovered for $m_0=0$ and $V_0=D\tau$.

The kernels introduced above can then be rewritten directly in terms of covariance projections. For example,
\beeq{
A^{(vv)}_{\omega,T}&=\int_{0}^{T} ds\left(\int_{s}^{T} dt \cos(\omega t) e^{-(t-s)/\tau}\right)^2\\
&=\int_{0}^{T} ds\left(\int_{s}^{T} dt \cos(\omega t) e^{-(t-s)/\tau}\right)\left(\int_{s}^{T} dt'\cos( \omega t') e^{-(t'-s)/\tau}\right)\\
&=\int_{0}^{T} ds\int_{s}^{T} dt\int_{s}^{T} dt' \cos(\omega t)\cos(\omega t') e^{-(t-s)/\tau}e^{-(t'-s)/\tau}\\
&=\int_{0}^{T} dt\int_{0}^{T} dt' \cos(\omega t)\cos(\omega t')\int_{0}^{\min(t,t')} ds e^{-(t+t'-2s)/\tau}\\
&=\frac{\tau}{2}\int_{0}^{T} dt\int_{0}^{T} dt' \cos(\omega t)\cos(\omega t')\Big(e^{-|t-t'|/\tau}-e^{-(t+t')/\tau}\Big)\,.
\label{eq:I_f_2D}
}
Using the expression of the two-time covariance, this becomes
\beeq{
A^{(vv)}_{\omega,T}=\frac{1}{2D}\int_{0}^{T} dt\int_{0}^{T} dt'  \cos(\omega t)\cos(\omega t')\Big\{C(t,t')-V_0 e^{-(t+t')/\tau}\Big\}\,.
}
The same rewriting applies to the kernels $A^{(ww)}_{\omega,T}$ and $A^{(vw)}_{\omega,T}$.

%%%%%%%%%%%%%%%%%%%%%%%%%%%%%%%%%%%%%%%%%%%%%%%%%%%
\section{Multi-frequency case}
\label{appB}
%%%%%%%%%%%%%%%%%%%%%%%%%%%%%%%%%%%%%%%%%%%%%%%%%%%
We now derive the multi-frequency MGF. The calculation follows the same steps as in the single-frequency case treated in Appendix~\ref{appA}. Starting from the definition,
\beeq{
\Phi_{\bm{\omega},T}(\bm{\lambda})&=\mathbb{E}_{\{X(t)\}}\left[e^{- \sum_{i=1}^L\lambda_i S(\omega_i,T )}\right]\\
&=\mathbb{E}_{\{X(t)\}}\left\{\exp\left[ -\sum_{i=1}^L\frac{\lambda_i }{T} \left(\int_0^T dt  \cos(\omega_it)X(t)\right)^2 -\sum_{i=1}^L\frac{\lambda_i }{T}\left( \int_0^T dt \sin(\omega_it)X(t)\right)^2\right]\right\}\\
&=\int\left[\prod_{i=1}^L\frac{dv_i dw_i}{2\pi}\right] \exp\left[-\sum_{i=1}^L\left(\frac{v^2_i}{2}+\frac{w^2_i}{2}\right)\right]\mathbb{E}_{\{X(t)\}}\exp\left[i \sum_{i=1}^L\sqrt{\frac{2\lambda_i }{T}} \int_0^T dt  \left(v_i\cos(\omega_it)+w_i\sin(\omega_it)\right)X(t) \right]\,.
}
Introducing
\beeq{
g(t|v_i,w_i,\omega_i)&\equiv v_i\cos(\omega_i t)+w_i\sin(\omega_i t)\,,\\
\mathcal{F}(s|v_i,w_i,\omega_i,T)&\equiv \int_s^T dt\,g(t|v_i,w_i,\omega_i)e^{-(t-s)/\tau}\,,
}
and using the formal solution of the OU process, the Gaussian average over the noise history yields
\beeq{
&\mathbb{E}_{\{W_x(t)\}}\exp\left[2i \sum_{i=1}^L\sqrt{\frac{\lambda_i D}{T}} \int_0^T ds W_x(s)\mathcal{F}(s|v_i,w_i,\omega_i,T) \right]\\
&=\frac{1}{\mathcal{Z}}\int \mathcal{D}W_x \exp\left[-\frac{1}{2}\int_0^T ds W_x^2(s)+2i \sum_{i=1}^L\sqrt{\frac{\lambda_i D}{T}} \int_0^T ds W_x(s)\mathcal{F}(s|v_i,w_i,\omega_i,T) \right]\\
&=\frac{1}{\mathcal{Z}}\int \mathcal{D}W_x \exp\left[-\frac{1}{2}\int_0^T ds \left(W_x(s)-2i\sum_{i=1}^L  \sqrt{\frac{\lambda_i D}{T}} \mathcal{F}(s|v_i,w_i,\omega_i,T)\right)^2
+\frac{1}{2}\int_0^Tds\left(2i\sum_{i=1}^L  \sqrt{\frac{\lambda_i D}{T}} \mathcal{F}(s|v_i,w_i,\omega_i,T)\right)^2
\right]\\
&= \exp\left[-\frac{2D}{T}\sum_{i,j=1}^L \sqrt{\lambda_i\lambda_j}  \int_0^Tds \mathcal{F}(s|v_i,w_i,\omega_i,T)\mathcal{F}(s|v_j,w_j,\omega_j,T)\right]\,.
}
Therefore,
\beeq{
\Phi_{\bm{\omega},T}(\bm{\lambda})&=\int\left[\prod_{i=1}^L\frac{dv_i dw_i}{2\pi}\right] \exp\left[-\sum_{i=1}^L\left(\frac{v^2_i}{2}+\frac{w^2_i}{2}\right)-\frac{2D}{T}\sum_{i,j=1}^L \sqrt{\lambda_i\lambda_j}  \int_0^Tds \mathcal{F}(s|v_i,w_i,\omega_i,T)\mathcal{F}(s|v_j,w_j,\omega_j,T)\right]\,.
}
We now write
\beeq{
\int_0^Tds \mathcal{F}(s|v_i,w_i,\omega_i,T)\mathcal{F}(s|v_j,w_j,\omega_j,T)&=A^{(v_iv_j)}_{\omega_i,\omega_j,T} v_i v_j+A^{(w_iw_j)}_{\omega_i,\omega_j,T} w_i w_j\\
&+ A^{(v_iw_j)}_{\omega_i,\omega_j,T} v_i w_j+A^{(w_iv_j)}_{\omega_i,\omega_j,T} w_i v_j\,,
}
where
\beeq{
A^{(v_iv_j)}_{\omega_i,\omega_j,T}&\equiv\int_0^Tds\left(\int_s^T dt \cos(\omega_i t)e^{-(t-s)/\tau}\right)\left(\int_s^T dt \cos(\omega_j t)e^{-(t-s)/\tau}\right)\,,\\
A^{(w_iw_j)}_{\omega_i,\omega_j,T}&\equiv\int_0^Tds\left(\int_s^T dt \sin(\omega_i t)e^{-(t-s)/\tau}\right)\left(\int_s^T dt \sin(\omega_j t)e^{-(t-s)/\tau}\right)\,,\\
A^{(v_iw_j)}_{\omega_i,\omega_j,T}&\equiv\int_0^Tds\left(\int_s^T dt \cos(\omega_i t)e^{-(t-s)/\tau}\right)\left(\int_s^T dt \sin(\omega_j t)e^{-(t-s)/\tau}\right)\,,\\
A^{(w_iv_j)}_{\omega_i,\omega_j,T}&\equiv\int_0^Tds\left(\int_s^T dt \sin(\omega_i t)e^{-(t-s)/\tau}\right)\left(\int_s^T dt \cos(\omega_j t)e^{-(t-s)/\tau}\right)\\
&=A^{(v_j w_i)}_{\omega_j,\omega_i,T}\,.
}
Thus,
\beeq{
\Phi_{\bm{\omega},T}(\bm{\lambda})&=\int\left[\prod_{i=1}^L\frac{dv_i dw_i}{2\pi}\right] \exp\Bigg[-\sum_{i=1}^L\left(\frac{v^2_i}{2}+\frac{w^2_i}{2}\right)\Bigg]\\
&\times\exp\Bigg[-\frac{2D}{T}\sum_{i,j=1}^L \sqrt{\lambda_i\lambda_j} \left(A^{(v_iv_j)}_{\omega_i,\omega_j,T} v_i v_j+A^{(w_iw_j)}_{\omega_i,\omega_j,T} w_i w_j+ A^{(v_iw_j)}_{\omega_i,\omega_j,T} v_i w_j+A^{(w_iv_j)}_{\omega_i,\omega_j,T} w_i v_j\right) \Bigg]\,.
\label{eq:MGFmultifrequency}
}
For the inverse Laplace transform it is convenient to eliminate the square roots from the non-Gaussian part by the change of variables
\beeq{
\sqrt{\lambda_i}v_i\to v_i\,,\qquad \sqrt{\lambda_i}w_i\to w_i\,,\qquad i=1,\ldots,L\,.
}
This gives
\beeq{
\Phi_{\bm{\omega},T}(\bm{\lambda})&=\int\left[\prod_{i=1}^L\frac{dv_i dw_i}{2\pi}\right] \frac{1}{\prod_{i=1}^L \lambda_i}\exp\Bigg[-\sum_{i=1}^L\frac{1}{\lambda_i}\left(\frac{v^2_i}{2}+\frac{w^2_i}{2}\right)\Bigg]\\
&\times\exp\Bigg[-\frac{2D}{T}\sum_{i,j=1}^L \left(A^{(v_iv_j)}_{\omega_i,\omega_j,T} v_i v_j+A^{(w_iw_j)}_{\omega_i,\omega_j,T} w_i w_j+ A^{(v_iw_j)}_{\omega_i,\omega_j,T} v_i w_j+A^{(w_iv_j)}_{\omega_i,\omega_j,T} w_i v_j\right) \Bigg]\,.
\label{eq:MGFmultifrequency_rescaled}
}
In this form the entire nontrivial dependence on the kernels is independent of $\bm{\lambda}$, while derivatives with respect to $\lambda_i$ act only on the anisotropic Gaussian factor.

For later use, let us define
\beeq{
Q_{ij}&\equiv A^{(v_iv_j)}_{\omega_i,\omega_j,T} v_i v_j+A^{(w_iw_j)}_{\omega_i,\omega_j,T} w_i w_j+ A^{(v_iw_j)}_{\omega_i,\omega_j,T} v_i w_j+A^{(w_iv_j)}_{\omega_i,\omega_j,T} w_i v_j\,.
}
Expanding Eq.~\eqref{eq:MGFmultifrequency} and performing the Gaussian averages over the independent standard normal variables $\{v_i,w_i\}$, one finds
\beeq{
\mathbb{E}\{S(\omega_i,T)\}&=\frac{2D}{T}\left\langle Q_{ii}\right\rangle_{\bm{v},\bm{w}}=\frac{2D}{T}\left(A^{(v_iv_i)}_{\omega_i,\omega_i,T} +A^{(w_iw_i)}_{\omega_i,\omega_i,T} \right)\,,\\
\mathbb{E}\{[S(\omega_i,T)]^2\}&=\left(\frac{2D}{T}\right)^2\left\langle Q_{ii}^2\right\rangle_{\bm{v},\bm{w}}\\
&=\left(\frac{2D}{T}\right)^2\Bigg\{3[A^{(v_iv_i)}_{\omega_i,\omega_i,T}]^2+3[A^{(w_iw_i)}_{\omega_i,\omega_i,T}]^2+4[A^{(v_iw_i)}_{\omega_i,\omega_i,T}]^2\\
&+2A^{(v_iv_i)}_{\omega_i,\omega_i,T} A^{(w_iw_i)}_{\omega_i,\omega_i,T}\Bigg\}\,,
}
and, for $i\neq j$,
\beeq{
\mathbb{E}\{S(\omega_i,T)S(\omega_j,T)\}&=\left(\frac{2D}{T}\right)^2\left(\left\langle Q_{ii}Q_{jj}\right\rangle_{\bm{v},\bm{w}}+\left\langle Q_{ij}Q_{ij}\right\rangle_{\bm{v},\bm{w}}+\left\langle Q_{ij}Q_{ji}\right\rangle_{\bm{v},\bm{w}}\right)\\
&=\left(\frac{2D}{T}\right)^2
\left(A^{(v_iv_i)}_{\omega_i,\omega_i,T} +A^{(w_iw_i)}_{\omega_i,\omega_i,T} \right)
\left(A^{(v_j v_j)}_{\omega_j,\omega_j,T}  +A^{(w_j w_j)}_{\omega_j,\omega_j,T} \right) \\
&+2\left(\frac{2D}{T}\right)^2\left([A^{(v_iv_j)}_{\omega_i,\omega_j,T}]^2+[A^{(w_iw_j)}_{\omega_i,\omega_j,T}]^2+[A^{(v_iw_j)}_{\omega_i,\omega_j,T}]^2+[A^{(w_iv_j)}_{\omega_i,\omega_j,T}]^2\right)\,.
}
Hence,
\beeq{
\mathbb{E}\{S(\omega_i,T)S(\omega_j,T)\}-\mathbb{E}\{S(\omega_i,T)\}\mathbb{E}\{S(\omega_j,T)\}=2\left(\frac{2D}{T}\right)^2\left([A^{(v_iv_j)}_{\omega_i,\omega_j,T}]^2+[A^{(w_iw_j)}_{\omega_i,\omega_j,T}]^2+[A^{(v_iw_j)}_{\omega_i,\omega_j,T}]^2+[A^{(w_iv_j)}_{\omega_i,\omega_j,T}]^2\right)\,,
}
which is valid for all $i$ and $j$.

We next derive the joint PDF of the collection of variables $\{S(\omega_i,T)\}_{i=1}^L$. Since the nontrivial $\bm{\lambda}$-dependence of Eq.~\eqref{eq:MGFmultifrequency_rescaled} is now factorized over $i$, we may perform the inverse Laplace transform frequency by frequency using the standard identity
\beeq{
\mathcal{L}^{-1}_{\lambda\to s}\left[\frac{1}{\lambda}\exp\left(-\frac{a}{\lambda}\right)\right]=J_0\left(2\sqrt{as}\right)\,.
}
With $a=(v_i^2+w_i^2)/2$, this yields
\beeq{
\rho_{\bm{\omega},T}(\bm{s})&=\int \left[\prod_{i=1}^L \frac{dv_i dw_i}{2\pi}\right]\exp\Bigg[-\frac{2D}{T}\sum_{i,j=1}^L \left(A^{(v_iv_j)}_{\omega_i,\omega_j,T} v_i v_j+A^{(w_iw_j)}_{\omega_i,\omega_j,T} w_i w_j+ A^{(v_iw_j)}_{\omega_i,\omega_j,T} v_i w_j+A^{(w_iv_j)}_{\omega_i,\omega_j,T} w_i v_j\right) \Bigg]\\
&\times\prod_{i=1}^L J_0\left(\sqrt{2s_i\left(v_i^2+w_i^2\right)}\right)\,.
\label{eq:bgi}
}
It is convenient to write this expression in matrix form. Introducing the $2L\times 2L$ block matrix
\beeq{
\mathcal{A}=\begin{pmatrix}
\bm{A}^{(vv)}&\bm{A}^{(vw)}\\
[\bm{A}^{(vw)}]^T&\bm{A}^{(ww)}
\end{pmatrix}\,,
\label{eq:Amatrix}
}
we may rewrite Eq.~\eqref{eq:bgi} as
\beeq{
\rho_{\bm{\omega},T}(\bm{s})&=\int \frac{d\bm{v} d\bm{w}}{(2\pi)^L}\exp\left[-\frac{2D}{T}\begin{pmatrix}\bm{v}^T&\bm{w}^T
\end{pmatrix}\mathcal{A}\begin{pmatrix}\bm{v}\\\bm{w}
\end{pmatrix}\right]\prod_{i=1}^L J_0\left(\sqrt{2s_i\left(v_i^2+w_i^2\right)}\right)\,.
}
Equation \eqref{eq:bgi} is the exact oscillatory Bessel-Gaussian representation of the joint density. In the next appendix we rewrite it in an explicitly positive form.

%%%%%%%%%%%%%%%%%%%%%%%%%%%%%%%%%%%%%
\section{Positivity of the Bessel--Gaussian Integral}
\label{appC}
%%%%%%%%%%%%%%%%%%%%%%%%%%%%%%%%%%%%%
We start from the exact oscillatory representation of the joint density obtained in Eq.~\eqref{eq:bgi},
\beeq{
\rho_{\bm{\omega},T}(\bm{s})&=\int \left[\prod_{i=1}^L \frac{dv_i dw_i}{2\pi}\right]\exp\Bigg[-\frac{2D}{T}\sum_{i,j=1}^L \left(A^{(v_iv_j)}_{\omega_i,\omega_j,T} v_i v_j+A^{(w_iw_j)}_{\omega_i,\omega_j,T} w_i w_j+ A^{(v_iw_j)}_{\omega_i,\omega_j,T} v_i w_j+A^{(w_iv_j)}_{\omega_i,\omega_j,T} w_i v_j\right) \Bigg]\\
&\times\prod_{i=1}^L J_0\left(\sqrt{2s_i\left(v_i^2+w_i^2\right)}\right)\,.
\label{eq:bgi_appC}
}
It is convenient to rewrite the quadratic form in matrix notation. Introducing the $2L\times 2L$ block matrix
\beeq{
\mathcal{A}=\begin{pmatrix}
    \bm{A}^{(vv)}&\bm{A}^{(vw)}\\
    \left[\bm{A}^{(vw)}\right]^T&\bm{A}^{(ww)}
\end{pmatrix}\,,
\label{eq:Amatrix_appC}
}
and the vector
\beeq{
\bm{z}\equiv\begin{pmatrix}\bm{v}\\\bm{w}\end{pmatrix}\,,
}
Eq.~\eqref{eq:bgi_appC} becomes
\beeq{
\rho_{\bm{\omega},T}(\bm{s})=\int \frac{d\bm{v} d\bm{w}}{(2\pi)^L}\exp\left[-\frac{2D}{T}\bm{z}^T\mathcal{A}\bm{z}\right]\prod_{i=1}^L J_0\left(\sqrt{2s_i\left(v_i^2+w_i^2\right)}\right)\,.
}
Now we use the standard integral representation
\beeq{
J_0\left(\sqrt{2s_i\left(v_i^2+w_i^2\right)}\right)=\frac{1}{2\pi}\int_0^{2\pi}d\phi_i\,\exp\left[i\sqrt{2s_i}\left(v_i\cos\phi_i+w_i\sin\phi_i\right)\right]\,.
}
Applying this identity for each $i$ yields
\beeq{
\rho_{\bm{\omega},T}(\bm{s})&=\int \frac{d^L\vec{\phi}}{(2\pi)^L}\int \frac{d\bm{v} d\bm{w}}{(2\pi)^L}\exp\left[-\frac{2D}{T}\bm{z}^T\mathcal{A}\bm{z}+i\bm{\eta}(\bm{s},\vec{\phi})^T\bm{z}\right]\,,
\label{eq:rho_phi_rep}
}
where
\beeq{
\bm{\eta}(\bm{s},\vec{\phi})\equiv \begin{pmatrix}
\sqrt{2s_1}\cos\phi_1\\
\vdots\\
\sqrt{2s_L}\cos\phi_L\\
\sqrt{2s_1}\sin\phi_1\\
\vdots\\
\sqrt{2s_L}\sin\phi_L
\end{pmatrix}\,.
}
The integral over $(\bm{v},\bm{w})$ is now Gaussian and can be done exactly:
\beeq{
\int d\bm{v} d\bm{w}\,\exp\left[-\frac{2D}{T}\bm{z}^T\mathcal{A}\bm{z}+i\bm{\eta}^T\bm{z}\right]
=\frac{\pi^L}{\sqrt{\det\left(\frac{2D}{T}\mathcal{A}\right)}}\exp\left[-\frac{T}{8D}\bm{\eta}^T\mathcal{A}^{-1}\bm{\eta}\right]\,.
}
Substituting this result into Eq.~\eqref{eq:rho_phi_rep}, we obtain
\beeq{
\rho_{\bm{\omega},T}(\bm{s}) = \left(\frac{T}{8 \pi D}\right)^L \frac{1}{\sqrt{\det\mathcal{A}}}\int d^L\vec{\phi}\,\exp\left[-\frac{T}{8D}\bm{\eta}(\bm{s},\vec{\phi})^T\mathcal{A}^{-1}\bm{\eta}(\bm{s},\vec{\phi})\right]\,.
\label{eq:rho_positive_matrix}
}
This already makes the positivity of the joint density manifest: since $\mathcal{A}$ is positive definite, so is $\mathcal{A}^{-1}$, and therefore the integrand in Eq.~\eqref{eq:rho_positive_matrix} is non-negative for every $\vec{\phi}$.

To write the exponent in component form, let us denote by
\beeq{
[\mathcal{A}^{-1}]^{(\xi_k\xi_l)}_{\omega_k,\omega_l,T}
}
the entry of $\mathcal{A}^{-1}$ coupling the component $\xi_k$ at frequency $\omega_k$ to the component $\xi_l$ at frequency $\omega_l$, with $\xi=v$ or $w$. Defining
\beeq{
A'^{(\xi_k \xi_l)}_{\omega_k,\omega_l,T}\equiv \frac{T}{4D}\sqrt{s_ks_l}\,[\mathcal{A}^{-1}]^{(\xi_k \xi_l)}_{\omega_k,\omega_l,T}\,,
}
we can expand the quadratic form as
\beeq{
\frac{T}{8D}\bm{\eta}(\bm{s},\vec{\phi})^T\mathcal{A}^{-1}\bm{\eta}(\bm{s},\vec{\phi})
=\sum_{k,l=1}^L\sum_{\sigma,\tau=0,1}A'^{(\xi_k \xi_l)}_{\omega_k,\omega_l,T}\cos(\phi_k-\sigma\pi/2)\,\cos(\phi_l-\tau\pi/2)\,,
}
where $\xi_k=v_k$ if $\sigma=0$ and $\xi_k=w_k$ if $\sigma=1$, and similarly for $\xi_l$. Therefore,
\beeq{
\rho_{\bm{\omega},T}(\bm{s}) = \left(\frac{T}{8 \pi D}\right)^L \frac{1}{\sqrt{\det\mathcal{A}}} \int d^L\vec{\phi}\,\exp\left[-\sum_{k,l=1}^L\sum_{\sigma,\tau=0,1}A'^{(\xi_k \xi_l)}_{\omega_k,\omega_l,T}\cos(\phi_k-\sigma\pi/2)\,\cos(\phi_l-\tau\pi/2)\right]\,.
\label{eq:rho_positive_final}
}
Equation \eqref{eq:rho_positive_final} is the desired explicitly positive representation of the Bessel--Gaussian integral.

%%%%%%%%%%%%%%%%%%%%%%%%%%%%%%%%%%%%%%%%%%%%%%%%%%%
\section{Multivariate Gaussian representation}
\label{appD}
%%%%%%%%%%%%%%%%%%%%%%%%%%%%%%%%%%%%%%%%%%%%%%%%%%%
We recall the cosine and sine projections defined in Eq.~\eqref{eq:definition_Z} on the observation window $[0,T]$,
\beeq{
Z_{c,i}\equiv \int_0^Tdt\cos(\omega_i t) X(t),\qquad Z_{s,i}\equiv \int_0^T dt \sin(\omega_i t)\,X(t),\qquad i=1,\dots,L\,.
}
Collecting them into the $2L$-dimensional vector
\beeq{
Z\equiv (Z_{c,1},\dots,Z_{c,L},Z_{s,1},\dots,Z_{s,L})^{T}\,,
}
we see immediately that $Z$ is multivariate Gaussian whenever the initial condition $X_0$ is Gaussian and independent of the driving noise, since the OU process is then Gaussian and $Z$ is obtained from it by linear operations. Therefore,
\beeq{
Z\sim \mathcal N(\bm{m}_Z,\Sigma)\,.
}
We now determine explicitly the mean vector $\bm{m}_Z$ and the covariance matrix $\Sigma$.

Recalling that $m_0=\mathbb{E}[X_0]$, $V_0=\text{Var}[X_0]$, and
\beeq{
m(t)\equiv \mathbb{E}[X(t)]=m_0 e^{-t/\tau}\,,
}
we obtain
\beeq{
\mathbb{E}[Z_{c,i}]&=\int_0^Tdt \cos(\omega_i t) m(t)\,,\\
\mathbb{E}[Z_{s,i}]&=\int_0^Tdt \sin(\omega_i t) m(t)\,.
}
It is convenient to define
\beeq{
u_c(\omega)&\equiv \int_0^T dt\,e^{-t/\tau}\cos(\omega t)=\frac{\tau}{1+(\omega\tau)^2}\Big(1-e^{-T/\tau}\cos(\omega T)+\omega\tau\,e^{-T/\tau}\sin(\omega T)\Big)\,,\\
u_s(\omega)&\equiv \int_0^T dt\,e^{-t/\tau}\sin(\omega t)=\frac{\tau}{1+(\omega\tau)^2}\Big(\omega\tau-e^{-T/\tau}\big(\omega\tau\cos(\omega T)+\sin(\omega T)\big)\Big)\,.
\label{eq:u_def}
}
Hence,
\beeq{
\mathbb{E}[Z_{c,i}] = m_0\,u_c(\omega_i),\qquad \mathbb{E}[Z_{s,i}] = m_0\,u_s(\omega_i)\,.
\label{eq:means_Z}
}
Introducing the vectors
\beeq{
\bm{u}_c\equiv \left(u_c(\omega_1),\dots,u_c(\omega_L)\right)^T,\qquad 
\bm{u}_s\equiv \left(u_s(\omega_1),\dots,u_s(\omega_L)\right)^T\,,
}
the mean vector can be written compactly as
\beeq{
\bm{m}_Z=m_0\begin{pmatrix}\bm{u}_c\\ \bm{u}_s\end{pmatrix}\,.
}

For the second moments, the covariances follow from the two-time covariance function
\beeq{
C(t,t')\equiv\mathbb{E}[X(t)X(t')]-\mathbb{E}[X(t)]\mathbb{E}[X(t')]=D\tau e^{-|t-t'|/\tau}+(V_0-D\tau)e^{-(t+t')/\tau}\,,
}
namely
\beeq{
\text{Cov}(Z_{c,i},Z_{c,j})&=\int_0^T dt\int_0^T dt'\;\cos(\omega_i t)\cos(\omega_j t')\,C(t,t')\,,\\
\text{Cov}(Z_{s,i},Z_{s,j})&=\int_0^T dt\int_0^T dt'\;\sin(\omega_i t)\sin(\omega_j t')\,C(t,t')\,,\\
\text{Cov}(Z_{c,i},Z_{s,j})&=\int_0^T dt\int_0^T dt'\;\cos(\omega_i t)\sin(\omega_j t')\,C(t,t')\,,\\
\text{Cov}(Z_{s,i},Z_{c,j})&=\int_0^T dt\int_0^T dt'\;\sin(\omega_i t)\cos(\omega_j t')\,C(t,t')\,.
\label{eq:Zcovariances}
}

We next relate these covariances to the kernels introduced in Eq.~\eqref{eq:A_multi_frequency}. The derivation is identical in structure to the single-frequency calculation of Appendix~\ref{appA}. For example,
\beeq{
A^{(v_iv_j)}_{\omega_i,\omega_j,T}&=\int_0^Tds\left(\int_s^T dt \cos(\omega_i t)e^{-(t-s)/\tau}\right)\left(\int_s^T dt' \cos(\omega_j t')e^{-(t'-s)/\tau}\right)\\
&=\int_0^T dt\int_0^T dt' \cos(\omega_i t)\cos(\omega_j t')\int_0^{\min(t,t')} ds e^{-(t+t'-2s)/\tau}\,.
}
Since
\beeq{
\int_0^{\min(t,t')} ds e^{-(t+t'-2s)/\tau}=\frac{\tau}{2}\Big(e^{-|t-t'|/\tau}-e^{-(t+t')/\tau}\Big)\,,
}
it follows that
\beeq{
A^{(v_iv_j)}_{\omega_i,\omega_j,T} =\frac{\tau}{2}\int_0^T dt\int_0^T dt' \cos(\omega_i t)\cos(\omega_j t')\Big(e^{-|t-t'|/\tau}-e^{-(t+t')/\tau}\Big)\,.
\label{eq:A_2D_kernel}
}
Exactly the same manipulation gives
\beeq{
A^{(w_iw_j)}_{\omega_i,\omega_j,T} &=\frac{\tau}{2}\int_0^T dt\int_0^T dt' \sin(\omega_i t)\sin(\omega_j t')\Big(e^{-|t-t'|/\tau}-e^{-(t+t')/\tau}\Big)\,,\\
A^{(v_iw_j)}_{\omega_i,\omega_j,T} &=\frac{\tau}{2}\int_0^T dt\int_0^T dt' \cos(\omega_i t)\sin(\omega_j t')\Big(e^{-|t-t'|/\tau}-e^{-(t+t')/\tau}\Big)\,,\\
A^{(w_iv_j)}_{\omega_i,\omega_j,T} &=\frac{\tau}{2}\int_0^T dt\int_0^T dt' \sin(\omega_i t)\cos(\omega_j t')\Big(e^{-|t-t'|/\tau}-e^{-(t+t')/\tau}\Big)\,.
}

Substituting the expression of $C(t,t')$ into Eqs.~\eqref{eq:Zcovariances}, we therefore obtain
\beeq{
\text{Cov}(Z_{c,i},Z_{c,j})&=2DA^{(v_iv_j)}_{\omega_i,\omega_j,T}+V_0 u_c(\omega_i)u_c(\omega_j)\,,\\
\text{Cov}(Z_{s,i},Z_{s,j})&=2DA^{(w_iw_j)}_{\omega_i,\omega_j,T}+V_0 u_s(\omega_i)u_s(\omega_j)\,,\\
\text{Cov}(Z_{c,i},Z_{s,j})&=2DA^{(v_iw_j)}_{\omega_i,\omega_j,T}+V_0 u_c(\omega_i)u_s(\omega_j)\,,\\
\text{Cov}(Z_{s,i},Z_{c,j})&=2DA^{(w_iv_j)}_{\omega_i,\omega_j,T}+V_0 u_s(\omega_i)u_c(\omega_j)\,.
\label{eq:Sigma_entries_general}
}

Writing these relations in block form, the covariance matrix reads
\beeq{
\Sigma=\begin{pmatrix}
2D\bm{A}^{(vv)}+V_0\bm{u}_c\bm{u}_c^T&
2D\bm{A}^{(vw)}+V_0\bm{u}_c\bm{u}_s^T\\
2D\left[\bm{A}^{(vw)}\right]^T+V_0\bm{u}_s\bm{u}_c^T&
2D\bm{A}^{(ww)}+V_0\bm{u}_s\bm{u}_s^T
\end{pmatrix}\,.
\label{eq:Sigma_general_appD}
}
This is the full multivariate Gaussian representation for arbitrary Gaussian initial condition.

In the main text we work with centered deterministic initialization, $m_0=0$ and $X_0=0$, for which $V_0=0$. In that case,
\beeq{
\bm{m}_Z=0\,,\qquad 
\Sigma=\begin{pmatrix}
2D\bm{A}^{(vv)}&2D\bm{A}^{(vw)}\\
2D\left[\bm{A}^{(vw)}\right]^T&2D\bm{A}^{(ww)}
\end{pmatrix}=2D\,\mathcal{A}\,,
\label{eq:Sigma_equals_A_appD}
}
where
\beeq{
\mathcal{A}=\begin{pmatrix}
    \bm{A}^{(vv)}&\bm{A}^{(vw)}\\
    \left[\bm{A}^{(vw)}\right]^T&\bm{A}^{(ww)}
\end{pmatrix}\,.
}
Therefore, under the assumptions used in Section~\ref{multiplefrequency},
\beeq{
Z\sim \mathcal N(0,\Sigma)\,,\qquad \Sigma=2D\mathcal{A}\,,
}
which is precisely the covariance-explicit lifted Gaussian representation used in the main text.

%%%%%%%%%%%%%%%%%%%%%%%%%%%%%%%%%%%%%%%%%%%%%%%%%%%
\section{Window mixing and the limits of window-only decorrelation}
\label{appE}
%%%%%%%%%%%%%%%%%%%%%%%%%%%%%%%%%%%%%%%%%%%%%%%%%%%
To clarify the discussion after Fig.~\ref{fig3}, it is useful to separate two statements that may be sometimes conflated. The first is that finite observation time mixes nearby Fourier components and thereby generates inter-frequency couplings on a scale of order $1/T$. The second, much stronger, statement would be that these couplings can be removed exactly from the data by using the rectangular window alone by, say, inverting the windowing operation itself. It is then useful to distinguish the window-induced mixing mechanism from the model-dependent covariance structure on which this mixing acts.

The finite-time covariance of the Fourier projections follows directly from Eq.~\eqref{eq:definition_Z}
\beeq{
\mathbb{E}_{\{X(t)\}}\left[\left(Z_{c,i}+iZ_{s,i}\right)\left(Z_{c,j}-iZ_{s,j}\right)\right]=\int_0^T\int_0^T dt dt' e^{i\omega_i t-i\omega_j t'}C(t,t')\,,
}
where $C(t,t')$ is the two-time covariance which has the following form for  centered deterministic initial conditions adopted in the main text
\beeq{
C(t,t')=D\tau\left(e^{-|t-t'|/\tau}-e^{-(t+t')/\tau}\right)\,.
}
For clarity of the discussion on the window-mixing mechanism, let us focus solely on the first term of the covariance, which corresponds to the stationary term:
\beeq{
C_{\rm st}(t-t')\equiv D\tau e^{-|t-t'|/\tau}\,.
}
Note that this stationary contribution can be written directly in terms of asymptotic standard PSD $\mu(\omega)$ introduced in Section~\ref{definitions},
\beeq{
C_{\rm st}(t)=\int_{-\infty}^\infty \frac{d\omega}{2\pi}\mu(\omega) e^{-i\omega t}\,.
}
This allows us to obtain the following:
\beeq{
\int_0^T dt\int _0^T dt' e^{i\omega_i t-i\omega_j t'}C_{\rm st}(t-t')=\int_{-\infty}^\infty \frac{d\omega}{2\pi}\mu(\omega) W_{T}(\omega_i-\omega)\overline{W_T(\omega_j-\omega)}\,,
\label{eq:window_overlap_appE}
}
with
\beeq{
W_T(\Omega)\equiv \int_0^T dt\,e^{i\Omega t}=e^{i\Omega T/2}\,\frac{2\sin(\Omega T/2)}{\Omega}\,.
}
Equation \eqref{eq:window_overlap_appE} makes the mechanism explicit: the rectangular window fixes the leakage scale $1/T$, while the spectral weight $\mu(\omega)$ sets the amplitude of the coupling. For the full OU covariance, the second term in $C(t,t')$ adds the finite-time preparation correction.  Hence, the exact inter-frequency couplings are not a function of the window alone: They depend on the process covariance itself and, through it, on the model parameters.

At the level of the mean spectrum, keeping only the stationary contribution reproduces the leakage formula
\beeq{
\frac{1}{T}\int_0^T dt\int_0^T dt'\,e^{i\omega (t-t')}C_{\rm st}(t-t')
=\frac{1}{T}\int_{-\infty}^{\infty}\frac{d\omega'}{2\pi}\,\mu(\omega')\,\left|W_T(\omega-\omega')\right|^2\,.
\label{eq:mean_leakage_appE}
}
Thus the stationary part of the finite-time mean spectrum is a smeared version of the asymptotic PSD. 

Understanding or correcting this leakage at the level of $\mathbb{E}_{\{X(t)\}}[S(\omega,T)]$ is, however, weaker than eliminating the joint fluctuations of the collection $\{S(\omega_i,T)\}_{i=1}^{L}$. The present work is concerned with this second problem.

There is also a basic information-theoretic obstruction. Multiplication by $w_T$ in time is a projection: it deletes the trajectory outside the observation interval $[0,T]$. Hence no exact inverse exists that reconstructs the unwindowed signal from the observed finite record alone. In frequency space this means that convolution with $W_T$ is not a benign blur with a universal stable inverse. The finite-$T$ inter-frequency couplings are therefore not a removable post-processing artifact; they are the spectral signature of having access only to a finite time interval.

For the Gaussian setting of Section~\ref{multiplefrequency}, one can nevertheless decorrelate the retained Fourier coefficients once the model is specified. Using Eq.~\eqref{eq:definition_Z} and Eq.~\eqref{eq:Sigma_equals_A}, we have
\beeq{
Z\sim \mathcal N(0,\Sigma)\,,\qquad \Sigma=2D\mathcal A\,.
}
Whenever $\Sigma$ is nonsingular, the transformed vector
\beeq{
\widetilde Z\equiv \Sigma^{-1/2}Z
}
has independent standard-normal components. This is an exact whitening of the retained-band Gaussian law. It does not, however, amount to a universal window correction because the matrix $\Sigma^{-1/2}$ depends on the full finite-$T$ covariance and hence on the model parameters through the kernels in Eq.~\eqref{eq:A_multi_frequency}. Moreover, the variables $\widetilde{Z}$ are linear combinations of all retained Fourier components and no longer correspond to spectral estimates at individual physical frequencies.

The same conclusion is visible directly from the exact multispectral Laplace transform \eqref{eq:Laplace_det_Z},
\beeq{
\Phi_{\bm{\omega},T}(\bm{\lambda})= \det\left(I+\frac{4D}{T}\mathcal A\Lambda\right)^{-1/2}\,.
}
Any exact factorization of this expression into independent modes is obtained only after diagonalizing $\mathcal A$, and therefore only after using the full model-dependent covariance structure itself. What the window argument provides is the scale and locality of the inter-frequency couplings. What the exact finite-$T$ theory provides is the covariance matrix that quantifies them. In this sense, the possibility of whitening for fixed parameters does not weaken the role of the present construction. On the contrary, it clarifies why $\mathcal A$ is the central object: it is precisely the information required either to work with the full correlated likelihood or to justify controlled blockwise approximations such as those used in Section~\ref{sec:inference}.

%%%%%%%%%%%%%%%%%%%%%%%%%%%%%%%%%%%%%%%%%%%%%%%%%%%
\section{Weighted multispectral powers and a factorization inequality}
\label{appF}
%%%%%%%%%%%%%%%%%%%%%%%%%%%%%%%%%%%%%%%%%%%%%%%%%%%
The total-power formulas quoted in Section~\ref{multiplefrequency} are recovered by the special choice $q_i\equiv 1$. A direct extension is obtained by considering the weighted multispectral power
\beeq{
Y_{\bm q}(\bm\omega,T)\equiv \sum_{i=1}^{L}q_i S(\omega_i,T)\,,\qquad q_i\geq 0\,,
\label{eq:Yq_def_appF}
}
together with the diagonal matrix
\beeq{
\mathcal Q_{\bm q}\equiv \mathrm{diag}(q_1,\ldots,q_L,q_1,\ldots,q_L)\,.
\label{eq:Qq_def_appF}
}
Setting $\lambda_i=tq_i$ in Eq.~\eqref{eq:Laplace_det_Z} gives
\beeq{
\Phi_{\bm q,\bm\omega,T}(t)\equiv \mathbb{E}_{\{X(t)\}}\left[e^{-tY_{\bm q}(\bm\omega,T)}\right] = \det\left(I+\frac{4Dt}{T}\mathcal A\,\mathcal Q_{\bm q}\right)^{-1/2}\,.
\label{eq:Yq_laplace_1_appF}
}
Using the Weinstein–Aronszajn identity $\det(I+AB)=\det(I+BA)$, this may be rewritten as
\beeq{
\Phi_{\bm q,\bm\omega,T}(t) = \det\left(I+\frac{4Dt}{T}B_{\bm{q}}\right)^{-1/2}\,,\qquad
B_{\bm{q}}\equiv \mathcal Q_{\bm q}^{1/2}\mathcal A\,\mathcal Q_{\bm{q}}^{1/2}\,.
\label{eq:Yq_laplace_2_appF}
}
Since $B_{\bm{q}}$ is symmetric positive semidefinite, let $\beta_1,\ldots,\beta_r$ denote its strictly positive eigenvalues, with
\beeq{
r\equiv \operatorname{rank}(B_{\bm{q}})\leq 2m_{\bm{q}}\,,\qquad m_{\bm{q}}\equiv \#\{i:q_i>0\}\,.
\label{eq:Yq_rank_appF}
}
Then
\beeq{
\Phi_{\bm q,\bm\omega,T}(t) = \prod_{\ell=1}^{r}\left(1+\frac{4Dt}{T}\beta_\ell\right)^{-1/2}\,.
\label{eq:Yq_laplace_eigs_appF}
}
Differentiating Eq.~\eqref{eq:Yq_laplace_eigs_appF} at $t=0$ yields
\beeq{
\mu_{\bm{q},\bm{\omega},T}=\frac{2D}{T}\mathrm{Tr}B_{\bm{q}}\,,\qquad \sigma_{\bm{q},\bm{\omega},T}^{2}=\frac{8D^2}{T^2}\mathrm{Tr}\left(B_{\bm{q}}^{2}\right)\,,
\label{eq:Yq_moments_appF}
}
and therefore
\beeq{
\gamma_{\bm{q},\bm{\omega},T}^{2} = 2\frac{\mathrm{Tr}\left(B_{\bm{q}}^{2}\right)}{\left(\mathrm{Tr}B_{\bm{q}}\right)^2} = 2\frac{\sum_{\ell=1}^{r}\beta_\ell^{2}}{\left(\sum_{\ell=1}^{r}\beta_\ell\right)^2}\,.
\label{eq:Yq_gamma_appF}
}
Introducing the effective rank
\beeq{
r_{\rm eff}(B_{\bm{q}})\equiv \frac{\left(\mathrm{Tr}B_{\bm{q}}\right)^2}{\mathrm{Tr}\left(B_{\bm{q}}^{2}\right)}\,,
\label{eq:reff_appF}
}
Eq.~\eqref{eq:Yq_gamma_appF} becomes
\beeq{
\gamma_{\bm{q},\bm{\omega},T}^{2}=\frac{2}{r_{\rm eff}(B_{\bm{q}})}\,.
\label{eq:Yq_reff_appF}
}
Because the eigenvalues of $B_{\bm q}$ are nonnegative,
\beeq{
\sum_{\ell=1}^{r}\beta_\ell^2\leq \left(\sum_{\ell=1}^{r}\beta_\ell\right)^2
\qquad\text{and}\qquad
\left(\sum_{\ell=1}^{r}\beta_\ell\right)^2\leq r\sum_{\ell=1}^{r}\beta_\ell^2\,,
}
so that
\beeq{
\frac{2}{r}\leq \gamma_{\bm q,\bm\omega,T}^{2}\leq 2\,.
\label{eq:Yq_rank_bound_appF}
}
Using Eq.~\eqref{eq:Yq_rank_appF}, one obtains the simpler bound
\beeq{
\frac{1}{m_{\bm{q}}}\leq \gamma_{\bm{q},\bm{\omega},T}^{2}\leq 2\,.
\label{eq:Yq_mq_bound_appF}
}
The band-power result quoted in the main text is recovered by setting $q_i\equiv 1$, in which case $B_{\bm{q}}=\mathcal{A}$, while if only one weight is nonzero Eq.~\eqref{eq:Yq_mq_bound_appF} reduces to the single-frequency bound of Section~\ref{onefrequency}. Thus the multifrequency coefficient of variation is controlled not only by the number of retained frequencies, but more sharply by the effective rank of the weighted multispectral covariance structure.

A second consequence of the determinant representation is a comparison between the exact multispectral law and the fully factorized exact single-frequency approximation. Let $\mathcal A_{ij}$ denote the $2\times 2$ block of $\mathcal A$ coupling the frequencies $\omega_i$ and $\omega_j$, and define
\beeq{
\Phi_{\rm fac}(\bm\lambda)\equiv \prod_{i=1}^{L}\det\left(I_2+\frac{4D\lambda_i}{T}\mathcal{A}_{ii}\right)^{-1/2}=\prod_{i=1}^{L}\Phi_{\omega_i,T}(\lambda_i)\,.
\label{eq:Phi_fac_appF}
}
For $\bm{\lambda}\geq 0$, the block matrix $I+\frac{4D}{T}\Lambda^{1/2}\mathcal{A}\Lambda^{1/2}$ is positive semidefinite, so the block Hadamard--Fischer inequality gives
\beeq{
\det\left(I+\frac{4D}{T}\Lambda^{1/2}\mathcal{A}\Lambda^{1/2}\right) \leq \prod_{i=1}^{L}\det\left(I_2+\frac{4D\lambda_i}{T}\mathcal{A}_{ii}\right)\,.
\label{eq:HF_appF}
}
Taking the power $-1/2$ yields
\beeq{
\Phi_{\bm{\omega},T}(\bm{\lambda})\geq \Phi_{\rm fac}(\bm{\lambda})=\prod_{i=1}^{L}\Phi_{\omega_i,T}(\lambda_i)\,.
\label{eq:Phi_fac_compare_appF}
}
Thus the exact multispectral Laplace transform dominates the product of the exact single-frequency transforms. In the present context, this provides a precise mathematical form to the statement that deleting the cross-frequency blocks of the finite-$T$ covariance exaggerates effective independence.

\end{document}